\documentclass[apj,numberedappendix,twocolappendix,tighten]{emulateapj}
\slugcomment{Accepted for publication in the Astrophysical Journal} 

\usepackage{apjfonts}

\usepackage{graphicx}
\usepackage[colorlinks=true,citecolor=blue,linkcolor=magenta,urlcolor=cyan]{hyperref}
\usepackage{bm}
\usepackage{CJK}
\usepackage{color}
\usepackage{enumitem}
\usepackage{amsmath}
\usepackage[perpage]{footmisc}
\usepackage{multirow}
\usepackage{booktabs}

\usepackage{etoolbox}
\makeatletter
\patchcmd{\NAT@citex}
  {\@citea\NAT@hyper@{%
     \NAT@nmfmt{\NAT@nm}%
     \hyper@natlinkbreak{\NAT@aysep\NAT@spacechar}{\@citeb\@extra@b@citeb}%
     \NAT@date}}
  {\@citea\NAT@nmfmt{\NAT@nm}%
   \NAT@aysep\NAT@spacechar\NAT@hyper@{\NAT@date}}{}{}
\patchcmd{\NAT@citex}
  {\@citea\NAT@hyper@{%
     \NAT@nmfmt{\NAT@nm}%
     \hyper@natlinkbreak{\NAT@spacechar\NAT@@open\if*#1*\else#1\NAT@spacechar\fi}%
       {\@citeb\@extra@b@citeb}%
     \NAT@date}}
  {\@citea\NAT@nmfmt{\NAT@nm}%
   \NAT@spacechar\NAT@@open\if*#1*\else#1\NAT@spacechar\fi\NAT@hyper@{\NAT@date}}
  {}{}
\makeatother

\shorttitle{Spine Loops and EUV Late Phase}
\shortauthors{Sun et al.}

\begin{document}

\begin{CJK}{UTF8}{}

\title{
Hot Spine Loops and the Nature of a Late-Phase Solar Flare
}

\author{
\begin{CJK}{UTF8}{gbsn} 
Xudong Sun (孙旭东)\altaffilmark{1}, J. Todd Hoeksema\altaffilmark{1}, Yang Liu (刘扬)\altaffilmark{1}, Guillaume Aulanier\altaffilmark{2}, \\
Yingna Su (宿英娜)\altaffilmark{3}, Iain G. Hannah\altaffilmark{4}, Rachel A. Hock\altaffilmark{5}
\end{CJK}
}

\email{xudong@sun.stanford.edu}
\altaffiltext{1}{W. W. Hansen Experimental Physics Laboratory, Stanford University, Stanford, CA 94305, USA}
\altaffiltext{2}{LESIA, Observatoire de Paris, CNRS, UPMC, Univ. Paris Diderot, 5 place Jules Janssen, 92190 Meudon, France}
\altaffiltext{3}{Harvard-Smithsonian Center for Astrophysics, Cambridge, MA 02138, USA}
\altaffiltext{4}{SUPA School of Physics and Astronomy, University of Glasgow, Glasgow, G12 8QQ, UK}
\altaffiltext{5}{Space Vehicle Directorate, Air Force Research Laboratory, Kirtland Air Force Base, NM 87116, USA}


\begin{abstract}
The fan-spine magnetic topology is believed to be responsible for many curious features in solar explosive events. A spine field line links distinct flux domains, but direct observation of such feature has been rare. Here we report a unique event observed by the \textit{Solar Dynamic Observatory} where a set of hot coronal loops (over 10~MK) connected to a quasi-circular chromospheric ribbon at one end and a remote brightening at the other. Magnetic field extrapolation suggests these loops are partly tracer of the evolving spine field line. Continuous slipping- and null-point-type reconnections were likely at work, energizing the loop plasma and transferring magnetic flux within and across the fan quasi-separatrix layer. We argue that the initial reconnection is of the ``breakout'' type, which then transitioned to a more violent flare reconnection with an eruption from the fan dome. Significant magnetic field changes are expected and indeed ensued. This event also features an extreme-ultraviolet (EUV) late phase, i.e. a delayed secondary emission peak in warm EUV lines (about 2--7~MK). We show that this peak comes from the cooling of large post-reconnection loops beside and above the compact fan, a direct product of eruption in such topological settings. The long cooling time of the large arcades contributes to the long delay; additional heating may also be required. Our result demonstrates the critical nature of cross-scale magnetic coupling -- topological change in a sub-system may lead to explosions on a much larger scale.
\end{abstract}



\section{Introduction}
\label{sec:intro}

Magnetic topology plays a crucial role in solar explosive events, whether eruptive or confined. Topological structures (e.g. null, separatrix) outline distinctive flux domains, highlight discontinuities in magnetic connectivity, and serve as favorable sites for reconnection \citep[e.g.][]{priest2000}. In the classical scenario, field lines are brought into a diffusive region at the discontinuity by plasma inflow, where they instantly exchange footpoints and thus connectivity. This ``break-and-paste'' process transports magnetic flux across the topological boundaries. The field configuration relaxes and releases part of the excessive magnetic energy to power flares and coronal mass ejections \citep[CMEs; for reviews, see e.g.][]{forbes2006,hudson2011}. Topological structures may collectively characterize the coupling between multiple magnetic sub-systems. Changes in one active region (AR) or filament channel can quickly destabilize another despite their large spatial separation or size contrast, as demonstrated in the case of global-scale sympathetic eruptions \citep{schrijver2011,torok2011,titov2012}.

Locations where magnetic connectivity is continuous but with large gradient are known as quasi-separatrix layers \citep[QSLs;][]{priest1995}. Similar to separatrices (infinite gradient), they too tend to harbor hight electric current density \citep{aulanier2005}. The QSL footprints have been found to coincide with chromospheric flare ribbons \citep[e.g.][]{demoulin1997}, an evidence that reconnection can take place there in the absence of discontinuity. The connectivity exchange rate is finite owing to the finite field mapping gradient. Field lines may continuously ``slip'' within QSLs to exchange their footpoints, an extension of instantaneous break-and-paste scenario \citep{aulanier2006,aulanier2007}.

\begin{figure*}[th]
\centerline{\includegraphics{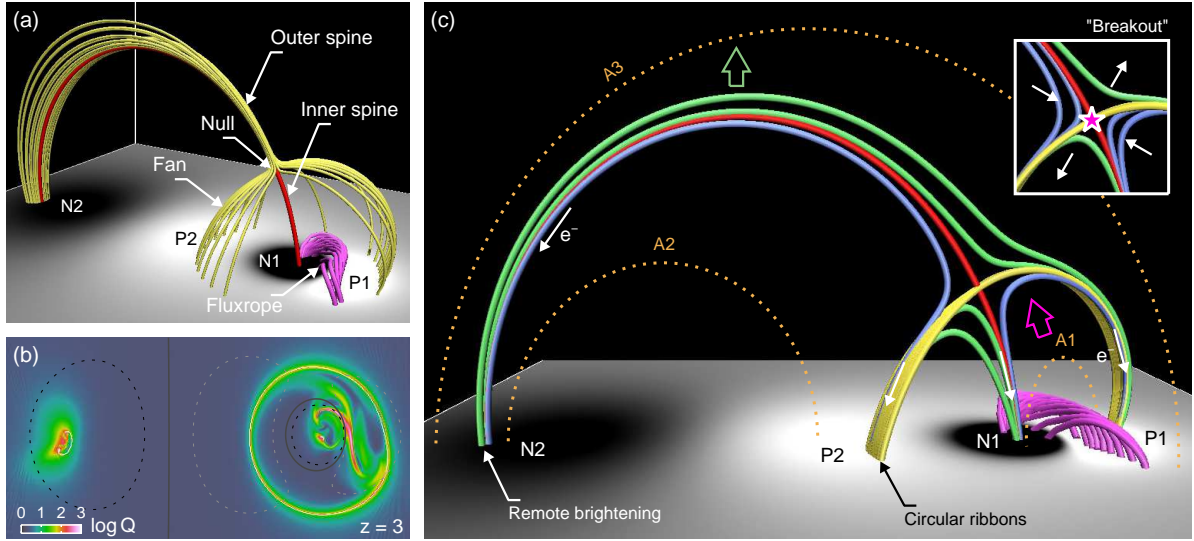}}
\caption{A nonlinear force-free field (NLFFF) toy model for explosion in fan-spine topology. (a) Magnetic skeleton of the model field. Four flux patches exist at the lower boundary (P1/N1 and P2/N2, P for positive, N for negative). A fluxrope (pink) resides above the P1/N1 polarity inversion line (PIL). The domain contains a null point above N1. The spine (red) and the fan (yellow) are marked. (b) Logarithm squashing degree $Q$ on the plane $z=3$. Red and yellow delineate the quasi-separatrix layer (QSL) footprint. Dotted lines are contours of the normal magnetic field; gray solid lines are the PILs. (c) Two-dimensional representation of the magnetic system illustrating possible physical processes pertinent to the observed event. Field lines undergo slipping-type reconnection within the fan QSL (yellow). At the null, blue field lines undergo breakout-type reconnection (inset). Energized particles or intense heat flux propagate along the fan and spine QSLs, resulting in circular ribbon and remote chromospheric brightening. The fluxrope may eventually erupt and open the fan. Post-reconnection loops of various configurations (A1, A2, and A3) may form.}
\label{f:cartoon}
\end{figure*}

Of particular interest here is the fan-spine topology \citep{lau1990} which often arises when new flux emerges into pre-existing field (see illustration in Figure~\ref{f:cartoon}(a)). One component of the new field becomes parasitic and surrounded by the opposite polarity. A null point subsequently forms in the corona \citep{torok2009}. From the null, a dome-shaped fan separatrix divides the two distinctive flux domains, below and above; a spine field line passes through both. 


Explosion in fan-spine topology yields interesting emission signatures. Some jetting ARs, for example, display anemone-like loops \citep{shibata1994} that outline the fan dome. Erupting and down-falling plasmas trace the open spine field line and the fan \citep{liuw2011}. Another example is the quasi-circular ribbon flare \citep{masson2009,suyn2009,wanghm2012}. Heat flux and energized particles from reconnection near a null flow along the separatrices/QSLs, lighting up their footprints in the lower atmosphere. The dome-shaped fan creates the quasi-circular ribbon, and the spine maps to a remote brightening. The ribbon brightens sequentially, suggesting slipping-type reconnection in the fan QSL. Here, the reconnecting loops may provide decisive evidence for the fan-spine topology and the conjectured physical processes, but direct, unambiguous observation has been elusive \citep[although, as suggested in][]{aulanier2000,gulielmino2010,deng2013}.

Magnetic topology may also affect flare irradiance evolution. An interesting secondary irradiance peak was recently identified in ``warm'' extreme-ultraviolet (EUV) emission lines (e.g. \ion{Fe}{16} 335~{\AA}, temperature about 3~MK) as much as 1--2 hours after the main impulsive phase \citep{woods2011}, without apparent corresponding X-ray signatures. Interestingly, this ``late-phase'' emission does not come from the post-reconnection loops at the main flaring site, but rather a different, longer set of loops in the same region dubbed ``late-phase arcades''. For the delayed emission peak, recent studies have explored the possibility of additional heating \citep{hock2012}, as well as the long plasma cooling time scale \citep{liuk2013}. They suggested the interaction between neighboring magnetic systems as the underpinning, although a detailed magnetic topology analysis is still pending.

Here, we report a unique event with a M2-class flare and CME (\texttt{SOL2011-11-15T12:43L338C109}) that features \textit{both} the spine-fan topology and a clear EUV late phase. New observations include, in particular, a rarely observed hot loop bundle (over 10~MK) in EUV that traces the spine field line and co-evolves with the circular flare ribbon. The subsequent eruption leads to large post-reconnection loops that produce the late-phase emission. We specifically address the following topics:
\begin{enumerate}[noitemsep,topsep=0pt,parsep=0pt,partopsep=0pt,label=\textit{\alph*}),leftmargin=*]
\item the formation and dynamics of the spine loops;
\item the cause and consequence of the eruption;
\item the formation and thermal evolution of the late-phase arcades.
\end{enumerate}

This event is well observed by the three-instrument suite on board the \textit{Solar Dynamic Observatory} \citep[\textit{SDO};][]{pesnell2012}. The Atmospheric Imaging Assembly \citep[AIA;][]{lemen2012} images the Sun in ten UV/EUV passbands. Its high cadence (12~s) and wide temperature coverage ($\log T$ between 3.7 and 7.3) are ideal for plasma morphology and thermal property analysis. The EUV Variability Experiment \citep[EVE;][]{woods2012} measures the solar EUV irradiance with moderate spectral resolution (1~{\AA}) and rapid cadence (10~s). The temporal evolution of the spectra helps identify the late phase and discern contributions of different lines in the same imaging passband. The Helioseismic and Magnetic Imager \citep[HMI;][]{schou2012} observes the spectropolarimetry of the \ion{Fe}{1} 6173~{\AA} absorption line. Full disk vector magnetograms are derived at a high cadence (12-minute) and moderate resolution (1$\arcsec$). Data characterization and reduction procedures are described in \cite{hoeksema2013}; a brief summary is given in the Appendix. The synergy between these state-of-the-art observations allows a comprehensive view of the event.

The paper is organized as follows. Section~\ref{sec:model} highlights several aspects of the explosions in the fan-spine topology based on literature. The key ingredients pertinent to this event are summarized in a toy model (Figure~\ref{f:cartoon}). In Sections~\ref{sec:spine} through~\ref{sec:latephase}, we present the event in detail and interpret the new observations in the framework of Section~\ref{sec:model}. The three sections deal with three stages of the evolution, which loosely correspond to the three topics mentioned above. We summarize our findings in Section~\ref{sec:conclude}. The Appendices document some of the analysis techniques and observational details. 


\section{Explosions in Fan-Spine Topology}
\label{sec:model}

\begin{figure*}[t]
\centerline{\includegraphics{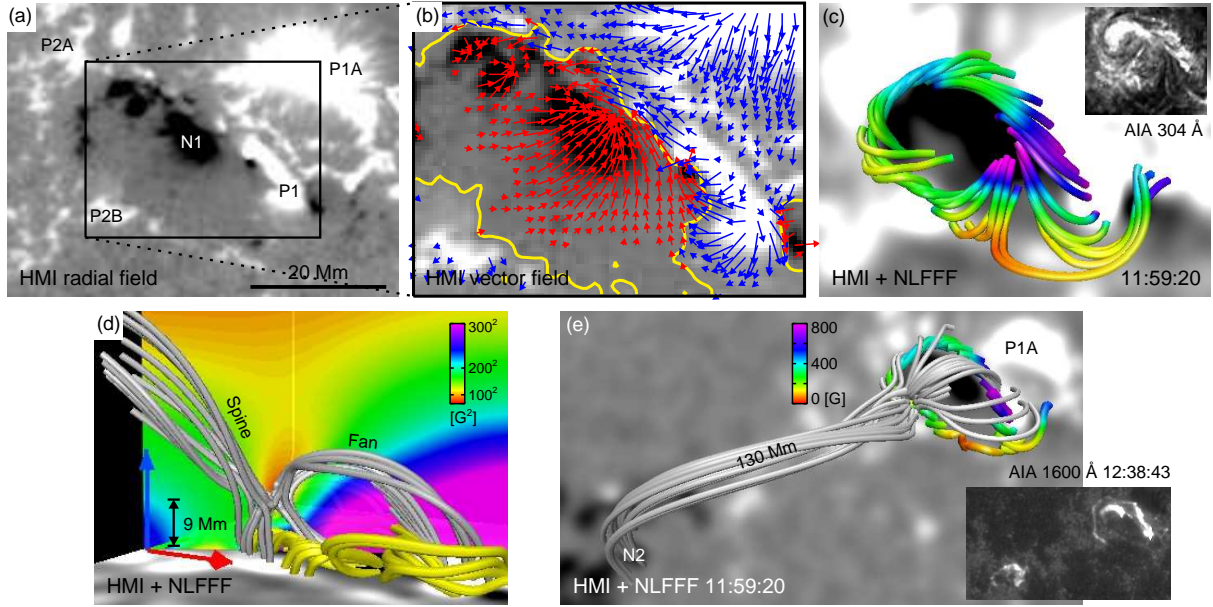}}
\caption{Magnetic field and topology of the pre-explosion AR at about 12:00 UT on 2011 November 15. (a) HMI radial field ($B_r$) map of the main flaring site. The field of view (FOV) is 58$\times$43~Mm$^2$. $B_r$ is scaled between $\pm$800~G. A small bipole (AR 11351, termed P1/N1) emerged inside the pre-existing positive polarity region (AR 11346). The old sunspot on the northwest is marked as P1A; the plage regions on the east as P2A and P2B. This denotation is chosen to be consistent with the toy model (cf. Figure~\ref{f:cartoon}). (b) HMI vector field map. On the background $B_r$ grayscale image, blue/red arrows show the horizontal component with positive/negative $B_r$. The arrow length is proportional to its strength $B_h$; its direction shows azimuth. The shortest arrows correspond to $B_h=200$~G. Yellow curves are the smoothed PILs. (c) Low-lying field lines from the NLFFF extrapolation. They are traced from the $B_z=-10$~G contour at a $z\approx3$~Mm. Colors indicate field strength (color table in (e)). A co-temporal AIA 304~{\AA} image depicting an AR filament is shown for comparison. (d) Side view of the magnetic skeleton. Colors on the vertical cross section show $B^2$. A weak field region, rendered as red, situates at $z\approx9$~Mm and contains multiple null points. Field lines traced from around these nulls outline a dome-shaped fan surface and the spine. Low-lying field lines in (c) are shown here in yellow. (e) Top-view of the magnetic skeleton. The spine field lines, with a typical length of 130~Mm, connect to the pre-existing negative flux marked as N2. This field line bundle displays apparent twist. An AIA 1600~{\AA} image of the flare ribbons is shown for context. All HMI maps are de-rotated to the disk center and remapped using a cylindrical equal-area projection.}
\label{f:field}
\end{figure*}

A fan-spine topology with a \textit{closed} field configuration may be constructed by two oppositely directed dipoles \citep[Figure~\ref{f:cartoon}(a); see also][]{antiochos1998}. One dipole emulates a larger, pre-existing AR (P2/N2); the other emulates a newly emerged, compact bipole (P1/N1). We introduce a fluxrope along the the P1/N1 polarity inversion line (PIL) to model the non-potential component \citep{ballegooijen2004}, which is often observed in the new field \citep[e.g.][]{leka1996,sun2012nlfff}. This results in a nonlinear force-free field (NLFFF) model, which is appropriate for the low plasma-$\beta$ corona. Its structure mimics the observed AR (Figure~\ref{f:field}). This toy model provides a clear template containing the essential features, well suited for illustration. Details of the toy model can be found in Appendix~\ref{a:model}.

Here, a single null point exists in the domain above the parasitic N1, through which a spine field line links N1 and N2. The fan separatrix divides the domain into two flux systems; it intersects the lower boundary to form a quasi-circular footprint. We calculate the squashing degree $Q$ on the lower boundary (Figure~\ref{f:cartoon}(b)) which measures the gradient of field line mapping \citep{demoulin1996,titov2002}. High-$Q$ regions correspond to the separatrices/QSLs. They include the footprint of the fan dome, the inner and outer spine, and an inverse-S-shaped pattern due to the left-hand-twisted fluxrope \citep[e.g.][]{savcheva2012}.

Numerical simulations show that flux emergence and line-tied boundary motion both can induce current sheet near a three-dimensional (3D) null point \citep[e.g.][]{pontin2007,moreno2008,pariat2009,edmondson2010}. The flux systems below and above the fan are then able to interact near the null through reconnection. Illustrated in our model, the blue field lines reconnect and become the green ones (inset of Figure~\ref{f:cartoon}(c)). Part of the P1 flux that originally links to N1 above the the fluxrope is transported out of the fan dome; it now maps to N2. Simultaneously, part of the P2 flux now connects to N1 instead of N2. In realistic cases, these topological structures may reside in more extended QSLs. Slipping-type reconnection may occur, transporting field lines within the QSLs toward and away from the null.

\textit{Hot spine loops.--} Reconnection-driven particles precipitate down the evolving fan and spine QSLs, resulting in a quasi-circular ribbon and remote brightening \citep{masson2009}; intense heat flux may do the same. Both processes can fill the loop with hot plasma through chromospheric evaporation. The temperature of the newly reconnected P1/N2 (green) loops, for example, may rise to over 10~MK, typical for the flaring loops. These loops will flow along the spine and outer part of the fan, thus trace out their evolution.

Reports of such hot topological portrait were rare prior to \textit{SDO}. Early soft X-ray (SXR) telescopes perhaps do not have the necessary spatial and temporal resolution, although hints were given in ``giant arch'' observations \citep[e.g.][]{simberova1993}. The \textit{Hinode} XRT telescope \citep{golub2007} has the required capability, but we are not aware of such studies using its observations. Previous EUV telescopes (EIT, TRACE, and EUVI), on the other hand, lack the passbands that are sensitive to higher temperatures.

The null-point reconnection in this setting is likely of the ``breakout'' nature \citep{antiochos1999}, that is, the reconnection involving weaker overlying field in a multipolar system. This process is generally mild, but is crucial to the subsequent eruptions as it reduces the magnetic tension force that stabilizes the system. Here, the null-point reconnection operates in a closed-field environment. The total amount of overlying flux is not reduced \citep{antiochos1998}; the process is largely \textit{confined} to begin with \citep[e.g.][]{masson2009}. Nevertheless, it does transports the immediate overlying field (P1/N1) out of the fan dome (P1/N2). This alleviates the constraints on the fluxrope; less energy is required to open the higher-lying loops.

We note that the eruption here originates from the ``outer lobe'' of the 3D fan dome (P1/N1 in Figure~\ref{f:cartoon}(c)), which renders the system highly asymmetric \citep{aulanier2000,mandrini2006}. This is somewhat different from the original, azimuthally invariant (i.e. 2.5-dimensional) breakout model which imposes shear on the ``central lobe'' \citep[P2/N1, e.g.][]{antiochos1999,lynch2008}.

Subsequent eruption in a similar environment has indeed been observed \citep{aulanier2000} and reproduced in simulations \citep{lugaz2011,jiangcw2013} for largely closed fan-spine structure. During this stage the null-point reconnection intensifies, and the fan dome becomes open. The fluxrope field may reconnect with the overlying field. A current sheet forms behind the erupting flux rope, where the two legs of the overlying field reconnect. This ``flare reconnection'' is predicted by the two-dimensional (2D), standard flare model \citep[for a recent review, see e.g.][]{hudson2011}. Contrary to the facilitative breakout reconnection preceding the the eruption, the flare reconnection is much more violent, as suggested by simulation \citep{karpen2012,lynch2013}. It is responsible for dissipating most of the magnetic energy in the system.


\begin{figure*}[t]
\centerline{\includegraphics{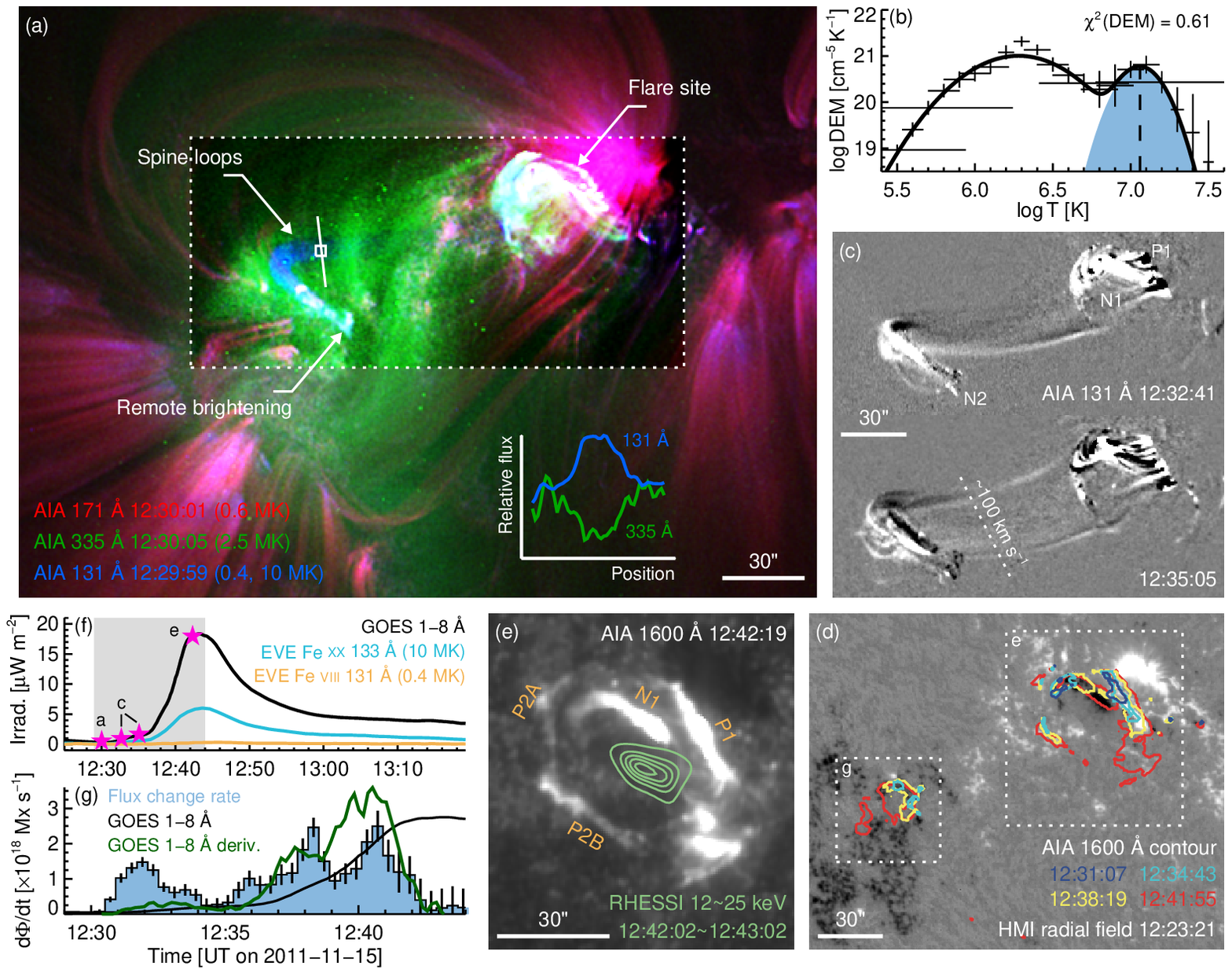}}
\caption{Characteristics of the spine loops and the flare ribbons. (a) Tricolor composite AIA image with central part enhanced. AIA passbands 171, 335, and 131~{\AA} are shown in red, green, and blue respectively. Sheared sigmoidal loops are bright at the flare site. Newly formed ``spine loops'' appear in the 131~{\AA} band connecting from the flaring site to a remote chromospheric brightening. Emission across the loop bundle (along white line segment) decreases in 335~{\AA} and increases in 131~{\AA} compared to the pre-flare state (see inset). (b) Regularized differential emission measure (DEM) inversion for a small loop segment (5$\times$5 pixel average, white square in (a)). A new peak appears at $\log T\approx7.1$, highlighted in cyan. The thick curve shows a double Gaussian fit in $\log T$ space. (c) AIA 131~{\AA} running difference images. The relevant photospheric field components are marked with P1, N1, and N2. The loop bundle expands in width at a speed of $\sim$100~km~s$^{-1}$ and shows apparent shuffling motion within. A two-frame average is applied before differencing. (d) Contours of AIA 1600~{\AA} (at 300~DN~s$^{-1}$ with pre-flare average subtracted) at four different times overplotted on a HMI radial magnetic field map (saturated at $\pm$1200~G) showing the evolution of the ribbons. The quasi-circular motions are obvious. The dotted boxes show the FOV for (e) and (g). (e) Flare ribbons in 1600~{\AA} near the flare peak. The overplotted contours are for \textit{RHESSI} 12--25~keV hard X-ray source (reconstructed with MEM NJIT algorithm, at 0.9, 0.7, 0.5, 0.3, and 0.1 of the maximum). Various segments of the ribbon are denoted according to the underlying photospheric field. The brightening at the southernmost shows apparent upward motion. (f) Light curve for \textit{GOES} 1--8~{\AA} soft X-ray, EVE \ion{Fe}{20}/\ion{Fe}{23} 133~{\AA}, and \ion{Fe}{8} 131~{\AA} lines. All have pre-flare average subtracted. The time stamps of (a), (c), and (e) are marked. The shaded area denotes the time range considered in (g). (g) Estimated magnetic flux change rate at the remote brightening site ($\rm{\dot{\Phi}_{N2}}$). The flux is calculated by summing all HMI pixels in (d) where AIA 1600~{\AA} count ever exceeded a 150~DN~s$^{-1}$ threshold (pre-flare subtracted). Error bars show the standard deviation from 11 trial computations where the threshold varies within $\pm10\%$. The \textit{GOES} light curve and its time derivative are also plotted in arbitrary units. (An animation of this figure is available online \href{http://sun.stanford.edu/~xudong/Article/Spine/corona.mp4}{at this url}.)}
\label{f:corona}
\end{figure*}


\textit{EUV late phase.--} The eruption yields multiple groups of post-reconnection loops with different connectivity. The strong null-point reconnection is expected to create mainly the A1- and A2-type (denoted in Figure~\ref{f:cartoon}(c)); the current-sheet reconnection the A1- and A3-type \citep{aulanier2000,lugaz2011}. If the fan is small compared to the pre-existing AR, the post-reconnection loops can be very different in size. They will cool at different rates during the initial, conduction-dominated stage, when the cooling time scales with the loop length squared \citep[for recent review, see][]{reale2010}. Because emission in the warm EUV lines increase only after the hot loops cool down, peaks from A2 and A3 loops will appear at a much later time compared to A1, as already noted in \cite{woods2011}. Additional heating from ongoing, weak reconnection may also contribute \citep{hock2012}. The two mechanisms need not be mutually exclusive.

Upon eruption, the horizontal component of the photospheric field generally displays a stepwise, irreversible increase near the main PIL P1/N1 \citep[e.g.][]{wanghm1994,sun2012nlfff,wangs2012}. Reconnection changes the coronal field connectivity, so the QSL footprints on the lower boundary are expected to change as well.


\section{The Hot Spine Loops}
\label{sec:spine}

NOAA AR 11346 is a decaying active region located in the southeastern quadrant of the solar disk on 2011 November 15. A new bipole, AR 11351, emerged within its leading polarity over one day prior. The negative polarity of the new bipole (denoted as N1) became the parasitic component (Figure~\ref{f:field}(a)). It was surrounded by the old, positive sunspot field (P1A) on the northwest, its own positive counterpart (P1) on the southwest, and the weaker, pre-existing positive field (P2A and P2B) on the eastern side. The negative component of AR 11346 (N2) was located over 100~Mm away on the southeast (Figure~\ref{f:field}(e)). The large spatial extent of the old AR is in sharp contrast with the compact new field.

The new bipole P1/N1 displayed clear shearing motion. We estimate their separation speed to be about 150~m~s$^{-1}$ by measuring the increasing distance between the P1 and N1 radial field ($B_r$) weighted centroids between 08:00 and 12:00 UT. Photospheric horizontal field (${\bf{B}}_h$) between P1/N1 is largely parallel to the PIL; the azimuth showed a clear rotational pattern near the northeast and southwest extremities (Figure~\ref{f:field}(b)). These indicate that the local magnetic field is highly non-potential. Rapid flux cancellation occurred between N1 and the surrounding positive field. The total unsigned flux in field of view (FOV) of Figure~\ref{f:field}(a) decreased by over 7\% between 00:00 and 12:00 UT. This is much greater than the $\sim$2\% daily variation caused by the \textit{SDO} spacecraft motion \citep{liuyang2012,hoeksema2013}, thus is likely to be real.

We perform a NLFFF extrapolation \citep{wiegelmann2004,wiegelmann2006} based on a series of HMI vector magnetograms. The frame at about 12:00 UT, roughly 20 minutes before the event, is shown here as an example. The modeling procedures are described in Appendix~\ref{a:topo} and are identical to those in \cite{sun2012nlfff}. Indeed, we find a low-lying, left-handed sheared field structure above the PIL (Figure~\ref{f:field}(c)). Its morphology agrees well with the inverse-S-shaped filament observed in the AIA \ion{He}{2} 304~{\AA} channel (thus dubbed the ``filament field''). Their strength can be as strong as 1000~G. The median torsional parameter measured at their footpoints, $\alpha=(\nabla \times {\bf{B}})_z/B_z$, is about $-0.3$~Mm$^{-1}$. The typical loop length $L$ is about 30~Mm. This yields an estimated twist of $|\alpha| L/2\approx4.5$ \citep{longcope2005}, about 0.7 turns.

Using a trilinear method \citep{haynes2007}, we find a group of null points clustered in a small weak field region about 9~Mm above N1 (rendered red in Figure~\ref{f:field}(d)). Field lines traced from this region flow downward and diverge to outline an inclined dome-shaped structure containing N1 and the filament field. Upward, they group into a bundle with some twist and map further away to N2 (Figure~\ref{f:field}(e)). Their typical length is about 130~Mm. The squashing degree $Q$ is indeed large (over 10$^{6}$) at the boundary outlined by these field lines and their footpoints (see Figure~\ref{f:qsl}). Some narrow bands of intermediate $Q$ values are also present. Very similar structures exist in the potential field (PF) extrapolation \citep[e.g.][]{sakurai1989}, where the topology is well defined. Details on the topology analysis can be found in Appendix~\ref{a:topo}. We therefore argue for the existence of fan-spine topology and extended QSLs. The magnetic structure resembles those reported in \cite{aulanier2000} and \cite{masson2009}.

The pre-explosion AR corona displays a temperature dichotomy (Figure~\ref{f:corona}(a)). Loops at the AR periphery were generally cooler. Those in the center that connect P2/N2 were warmer, showing a non-potential, forward-S shape. As early as 12:00 UT, AIA 1600~{\AA} images (flare emission dominated by \ion{C}{4}, from upper chromosphere to transition region) started to show sporadic brightening spots along the P1/N1 PIL. After 12:20 UT, they gradually developed into a pair of curved ribbons (P1/N1) that brightened sequentially. The front of the ribbons advanced clockwise along the filament (see Figures~\ref{f:corona}(d), (e)). Similar features are present in all AIA channels. The outer, P1 ribbon also advanced counterclockwise into P2A; it thus began to assume a quasi-circular shape around N1. A remote chromospheric brightening appeared in N2 with some counterclockwise rotational motion.

\begin{figure*}[th]
\centerline{\includegraphics{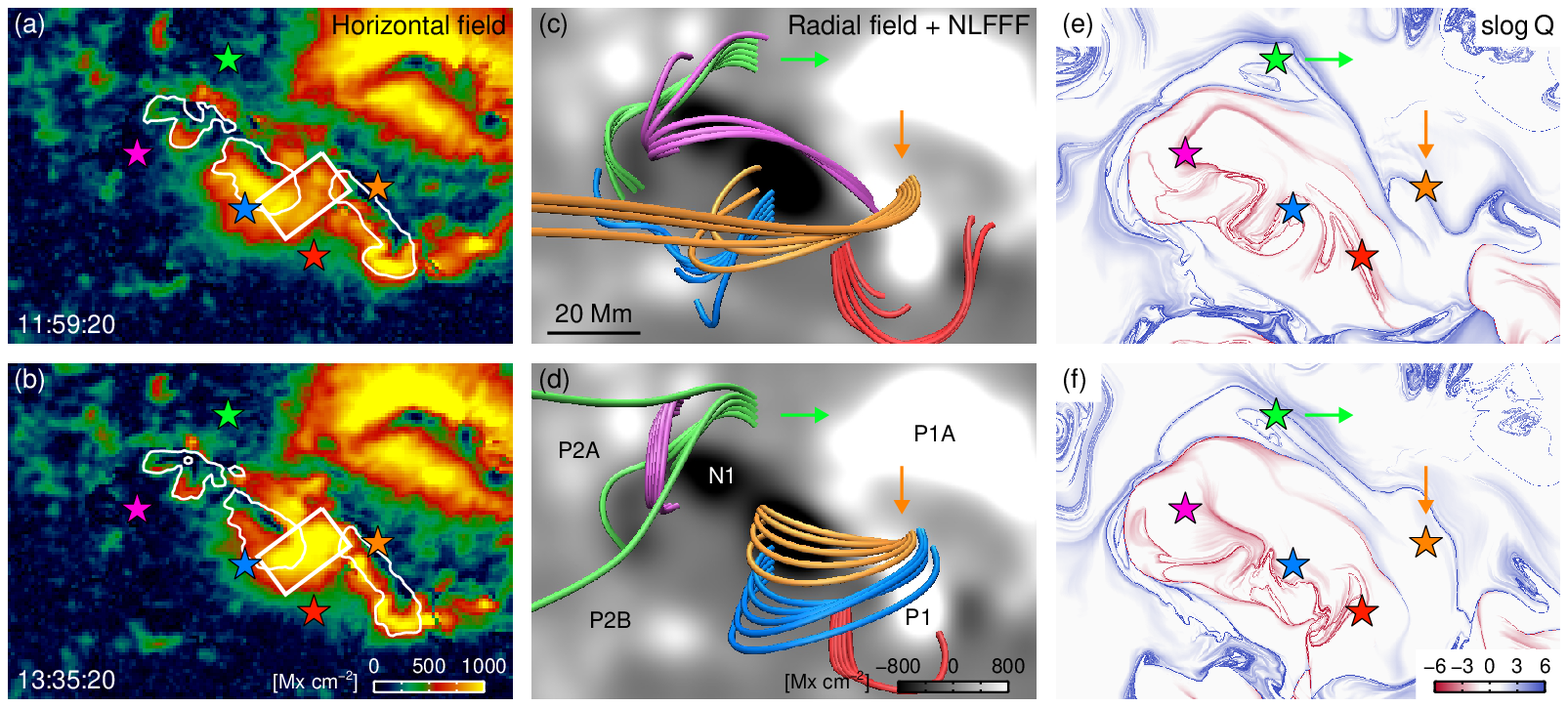}}
\caption{Magnetic field and connectivity changes due to the eruption. (a)-(b) Horizontal field ($B_h$) map before and after the eruption. The white contours outline the newly emerged bipole P1/N1 at $B_r=\pm400$~G. $B_h$ increases significantly between P1/N1 (boxed region), but decreases on the eastern edge of P1 (denoted by a blue star). (c)-(d) Selective NLFFF field lines from fixed footpoints, before and after. Five groups with different colors highlight the changes in magnetic connectivity. In particular, some green field lines open up or connect far away, while the orange ones close down. The blue field lines originally connected to P2A and P2B, now reach P1. The red and pink ones are shorter and less sheared. (e)-(f) Map of a signed logarithm version of the squashing degree (${\rm{slog}}\,Q$), before and after the event. Blue/red features denote large field line mapping gradient, i.e. the footprints of the QSLs, with positive/negative photospheric field. The star symbols are identical to (a) and (b), and denote the location of the field line footpoints in (c) and (d).}
\label{f:topo}
\end{figure*}

The geometry of the flare ribbons qualitatively matches the QSLs (or separatrices) derived from the NLFFF model (Figures~\ref{f:field}(e) and \ref{f:topo}(e)). For example, the circular ribbon and the remote brightening coincide with the fan and outer spine footprints. We note that the filament field might undergo internal, ``tether-cutting'' type reconnection \citep{moore2001}. The inner and outer footpoints (with respect to the PIL) of a sheared loop pair becomes connected respectively. The resultant sigmoidal loops connecting the outer footpoints were indeed observed in the SXR images taken by XRT, as well as in hot EUV passbands (131 and 94~{\AA}) early on. In this case, since the outer edge of the filament field is close to the northwestern segment of the fan, and inner edge close to the inner spine (Appendix~\ref{a:topo}), P1/N1 ribbons from the tether-cutting reconnection might be co-spatial with those from the evolving fan and the inner spine. The tether-cutting ribbons, however, are not expected to brighten sequentially. It is possible that both modes were at work, although a detailed investigation is out of the scope of this work. The ribbon evolutions are otherwise archetypical, as described in Section~\ref{sec:model}. 

We observe an interesting loop bundle that is most pronounced in the AIA 131~{\AA} passband (blue in Figure~\ref{f:corona}(a)) starting around 12:24 UT. Its western end mapped to the circular ribbon; its eastern end to the remote brightening (Figure~\ref{f:corona}(c)). The loop plasma is particularly hot. We perform a regularized differential emission measure (DEM) analysis \citep{hannah2012} on a small EUV loop segment using six AIA channels and find its temperature above 10~MK (Figure~\ref{f:corona}(b); see Appendix~\ref{a:dem} for details). This indicates that the increasing flux of the AIA 131~{\AA} band mainly comes from the hot \ion{Fe}{20}/\ion{Fe}{23} 133~{\AA} line rather than the cool \ion{Fe}{8} 131~{\AA}. This is confirmed by the EVE irradiance observation (Figure~\ref{f:corona}(f)). The simultaneous reduction of the 335~{\AA} flux (inset of Figure~\ref{f:corona}(a)) without accompanying eruption signatures suggests that these plasma was being heated from lower temperatures. These loops are also clearly visible in the SXR images from XRT.

\begin{figure*}[ht]
\centerline{\includegraphics{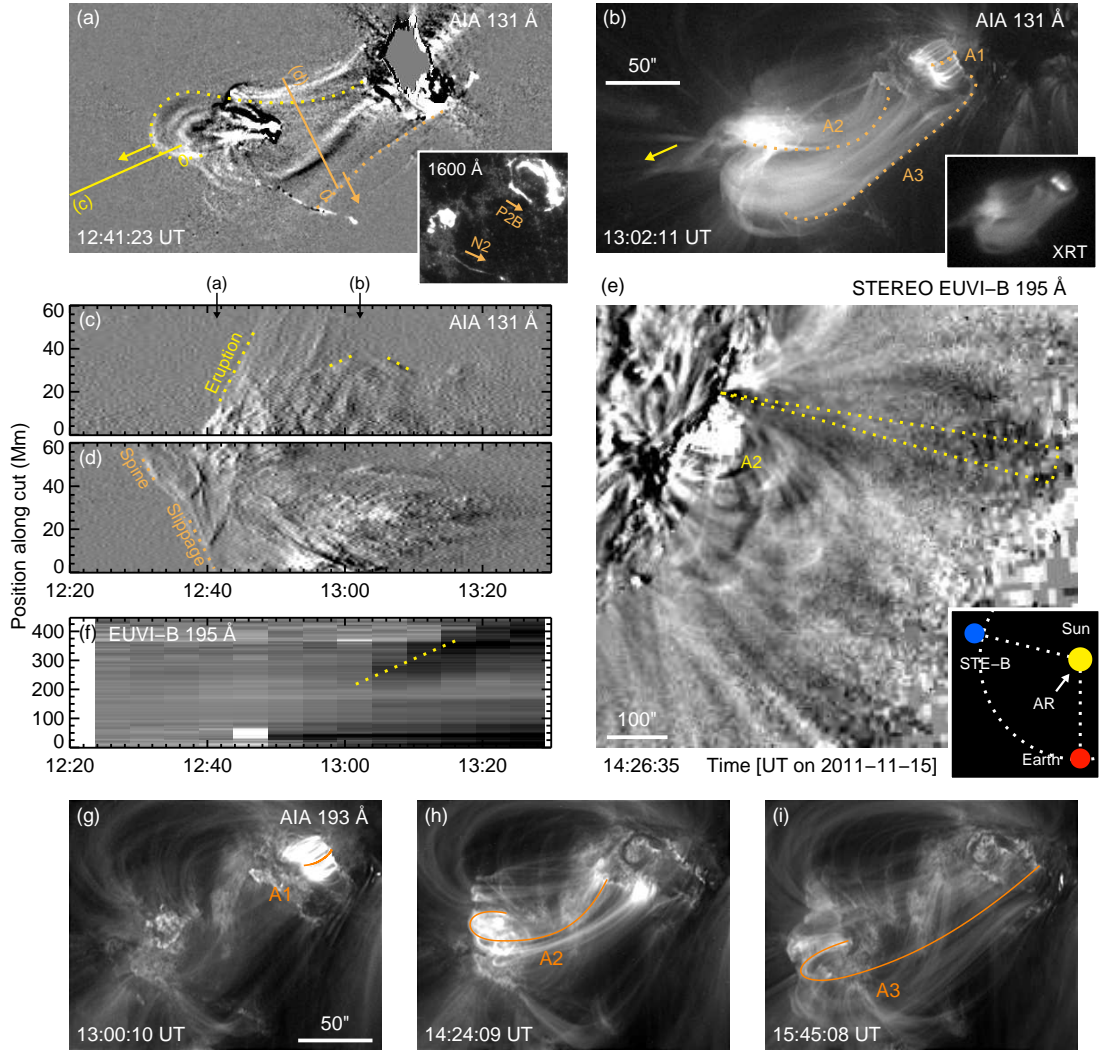}}
\caption{Morphology and kinematics of the coronal loops during and after the eruption. (a) AIA 131~{\AA} running difference image near the flare peak. The erupting loops are denoted by a yellow dotted curve; the slipping ones in orange. The inset shows a co-temporal 1600~{\AA} image. Two linear ribbons rapidly advance toward southwest. (b) AIA 131~{\AA} image 20 minutes after the flare. Multiple groups of post-reconnection loops are labeled (A1, A2, and A3; cf. Figure~\ref{f:cartoon}(c)). A cusp-shaped structure appears behind the ejecta (yellow arrow). The inset shows a co-temporal \textit{Hinode} XRT image with the Al-med filter. (c) Space-time diagram for the erupting loops, constructed by spatially sampling along the yellow solid line segment in (a) and stacking the samples in time. The eruption has a projected initial speed of about 150~km~s$^{-1}$. Some material moves up and falls back (two dotted lines on the right). (d) Similar to (c), but along the orange solid line segment in (a). The lateral motion and expansion of the spine loops have a speed of about 100~km~s$^{-1}$. The slipping loops move at a similar speed. Shuffling motions create interweaving tracks at a few tens of kilometers per second. (e) Base ratio 195~{\AA} image taken by \textit{STEREO}-B EUVI 1.8 hours after the flare. The eruption leaves behind a loop-like dimming region. Post-reconnection loops (A2-type) are also marked. The base image is taken at 12:12 UT. The ratio is scaled between 0.8 and 1.2. The relative locations of the Sun, the Earth, \textit{STEREO}-B, and the AR are shown in the inset. (f) Space-time diagram constructed from the narrow fan-shaped slit based at the fan dome in (e). The estimated speed of the ejecta front is approximately 170~km~s$^{-1}$. (g)-(i) AIA 193~{\AA} images. Three groups of post-reconnection loops with very different sizes and connectivities (A1, A2, and A3) show up about an hour apart. The A3 loops in (i) has the same connectivity as those A3 loops in (b), but are even longer and higher-lying. (An animation of this figure is available online \href{http://sun.stanford.edu/~xudong/Article/Spine/arcade.mp4}{at this url}.)}
\label{f:arcade}
\end{figure*}

Taking into account the magnetic structure, we interpret this loop bundle as the newly reconnected P1/N2 loops that flow along the evolving spine field line and part of the fan. We thus dub them the ``spine loops''. Propagating from the null, energetic particles and heat flux energized the chromosphere, which evaporated to fill these loops with hot plasma. As the event progressed, more loops should reconnect in this fashion, and the slipping-type reconnection should shuffle the footpoints of these loops in the fan QSL. Indeed, we observe the ribbons themselves grew in size. The loop bundle expanded in width at a rate of $\sim$100~km~s$^{-1}$ (Figures~\ref{f:corona}(c),~\ref{f:arcade}(d)). The western foot of the loop bundle's expanding edge appeared to link to the moving front of the circular ribbons. The intensity variations inside the loop bundle in the running difference images are suggestive of shuffling-like motions.

We use co-aligned HMI $B_r$ maps and AIA 1600~{\AA} images (Figure~\ref{f:corona}(d)) to estimate the total reconnected magnetic flux $\Phi$ and its change rate $\dot{\Phi}$ as a probe for the reconnection process \citep[e.g.][]{qiu2005,wanghm2012}. Magnetic flux in pixels swept by the flare ribbons are considered to have reconnected. They are then summed within each polarity, that is, $\rm{\Phi_P}$ for positive and $\rm{\Phi_N}$ for negative (including $\rm{\Phi_{N1}}$ for the circular ribbon and $\rm{\Phi_{N2}}$ for the remote brightening), respectively (Figure~\ref{f:corona}(g)). At 12:36 UT, the total reconnected flux is estimated to be $(\rm{\Phi_P} + \rm{\Phi_N})/2\approx7\times10^{20}$ Mx, with $\rm{\Phi_P} / \rm{\Phi_N}\approx1.5$, which is considered to be relatively balanced given the large systematic uncertainties \citep{qiu2005}. Between 12:30 UT and 12:36 UT, the Pearson correlation coefficient between 15 pairs of $\rm{\Phi_{N1}}$ and $\rm{\Phi_{N2}}$ is 0.98, with a fitted slope of about 0.88. This indicates that similar amounts of negative flux were reconnected at the main flare site and the remote brightening during this early stage. The co-evolution trend suggests an intimate physical linkage between the two flux systems \citep[e.g.][]{alexander2006}. The 1600~{\AA} images became severely saturated after 12:36 UT from intense emission, making such estimation difficult for the main flare site.

We note that all the phenomena described above occurred early in the event when the process remained confined. The time derivative of SXR flux measured by the \textit{GOES} satellite, which is a proxy for the energy release rate \citep[``Neupert effect'',][]{neupert1968}, started its rapid ascent only after 12:36 UT (Figure~\ref{f:corona}(g)). The SXR flux did not peak until 12:43 UT. This suggests the reconnection was relatively slow in the beginning, which is expected for the conjectured breakout process.



\section{The Eruption}
\label{sec:eruption}

As the SXR flux kept increasing, the remote brightening intensified and exhibited counterclockwise motion. The P1/N1 ribbon pair advanced clockwise rapidly and also became much brighter (Figure~\ref{f:corona}(e) and animation). Sigmoidal loops connecting P1/N1 were then observed in all EUV passbands. The circular ribbon also advanced counterclockwise into the weaker field of P2A. Its brightness was weaker than the P1/N1 counterpart.

Starting around 12:40 UT, a linear ribbon suddenly developed on the southeastern periphery of the main flaring site in P2B and rapidly advanced southwestward (Figure~\ref{f:corona}(e) and inset of Figure~\ref{f:arcade}(a)). Within a minute, a dimmer ribbon spanning almost 100~Mm developed south to the remote brightening site in the weaker field region of N2. These two ribbons were largely parallel; they were clearly linked by 131~{\AA} loops which appeared to be slipping toward southwest (Figure~\ref{f:arcade}(a)). We place a cut approximately perpendicular to the slipping loops and estimate its projected speed to be about 100~km~s$^{-1}$ using the space-time diagram technique (Figure~\ref{f:arcade}(d)). Shuffling motions in both directions were discernible in the diagram as streaks with positive and negative slopes; their speeds are typically a few tens of kilometers per second.

Around 12:40 UT, some of the hot loops appeared to expand eastward and finally erupted (Figure~\ref{f:arcade}(a)). We placed another cut along the direction of the eruption and estimate the projected speed to be about 150~km~s$^{-1}$ (Figure~\ref{f:arcade}(c)). The eruption lasted for about 10 minutes, some of the material fell back by the end. These erupting loops were also observable in cooler EUV passbands, but in the form of reduced emission. The stretched loops left behind a long-lasting cusp-shaped structure (Figure~\ref{f:arcade}(b)), which hints at the current-sheet reconnection behind the ejecta. The P1/N1 filament was still present after the eruption (e.g. Figure~\ref{f:arcade}(h)).

The SECCHI/EUVI instrument \citep{howard2008} on the \textit{STEREO}-B satellite was about 103$^\circ$ behind the Earth in Heliographic longitude. It observed this event in the 195~{\AA} passband with a 5-minute cadence. A loop-shaped structure with reduced emission appeared to erupt from the fan dome (Figure~\ref{f:arcade}(e) and accompanying animation). The space-time diagram from a base-ratio image sequence shows that the projected speed of its front is about 170~km~s$^{-1}$ (Figure~\ref{f:arcade}(f)). If it indeed originated from the fan dome, backward extrapolation in the diagram places the onset time at 12:41 UT, right before the SXR peak, which is consistent with the AIA 131~{\AA} observation. However, the uncertainty here can be large here owing to the lower cadence and the poor intensity contrast. The ejecta became discernible only after about 13:00 UT and 200~Mm above the limb.

The hard X-ray (HXR) telescope \textit{RHESSI} \citep{lin2002} started observing near the SXR peak. A clear HXR source appeared to be encircled by the circular ribbon (Figure~\ref{f:corona}(e)). Interestingly, it did not coincide with any chromospheric emission enhancement, as opposed to an earlier case where the HXR sources overlap with the circular ribbon \citep{reid2012}. This suggests the source was of coronal origin. Its proximity to the inferred null on the plane-of-sky provides evidence of strong flare reconnection above the original fan dome.

The 1600~{\AA} ribbons started to exhibit perpendicular motions around the SXR peak. For example, the P1/N1 ribbon pair started moving away from each other. The eastern half of the circular ribbon and the remote brightening site both expanded in width. This perhaps corresponds to the typical feature in 2D-like standard model, that is, more flux is brought into the current sheet and reconnects there, higher in the corona. The ribbons moved well beyond the original QSL boundary, perhaps indicating the opening of the fan separatrix \citep{aulanier2000}.

On the southern extremity of P1, 1600~{\AA} images show upward motion that might be related to the eruption. Small-scale jet-like events were also present nearby. They were located outside the fan dome, above patches of negative flux that were colliding with P1. Their relation to the main eruption is not clear.

These observations show that within 10--20 minutes, the relatively mild, confined event quickly transitioned into an eruptive one with intensified energy release. The waiting time (during which the breakout reconnection was at work) before the onset is compatible with the prediction from simulation \citep[tens of minutes;][]{karpen2012}. It fits well into the scenario proposed in Section~\ref{sec:model}.

We speculate that shearing and flux cancellation facilitated the expansion of the core field \citep[e.g.][]{amari2000,moore2010}. It also initiated the breakout reconnection, which reduced the restraining of the overlying field. At a critical point, ideal \citep[e.g.][]{torok2005,aulanier2010} or resistive \citep[e.g.][]{karpen2012} instability set in and caused the eruption. The detailed onset mechanism, of course, requires further investigation.

The conjectured breakout process here is fully 3D. Unlike the 2D case, the eruption does not require the opening of all the background field. Presumably, some of the overlying loops may simply be ``pushed aside'' by the pressure impulse, rather than reconnecting with the core field or in the current sheet behind the ejecta. Numerical test of such scenario process so far has started with an out-of-equilibrium flux rope with high twist \citep{lugaz2011}. It remains to be seen whether realistic shearing and flux cancellation can lead to eruption. 


\begin{figure*}[h]
\centerline{\includegraphics{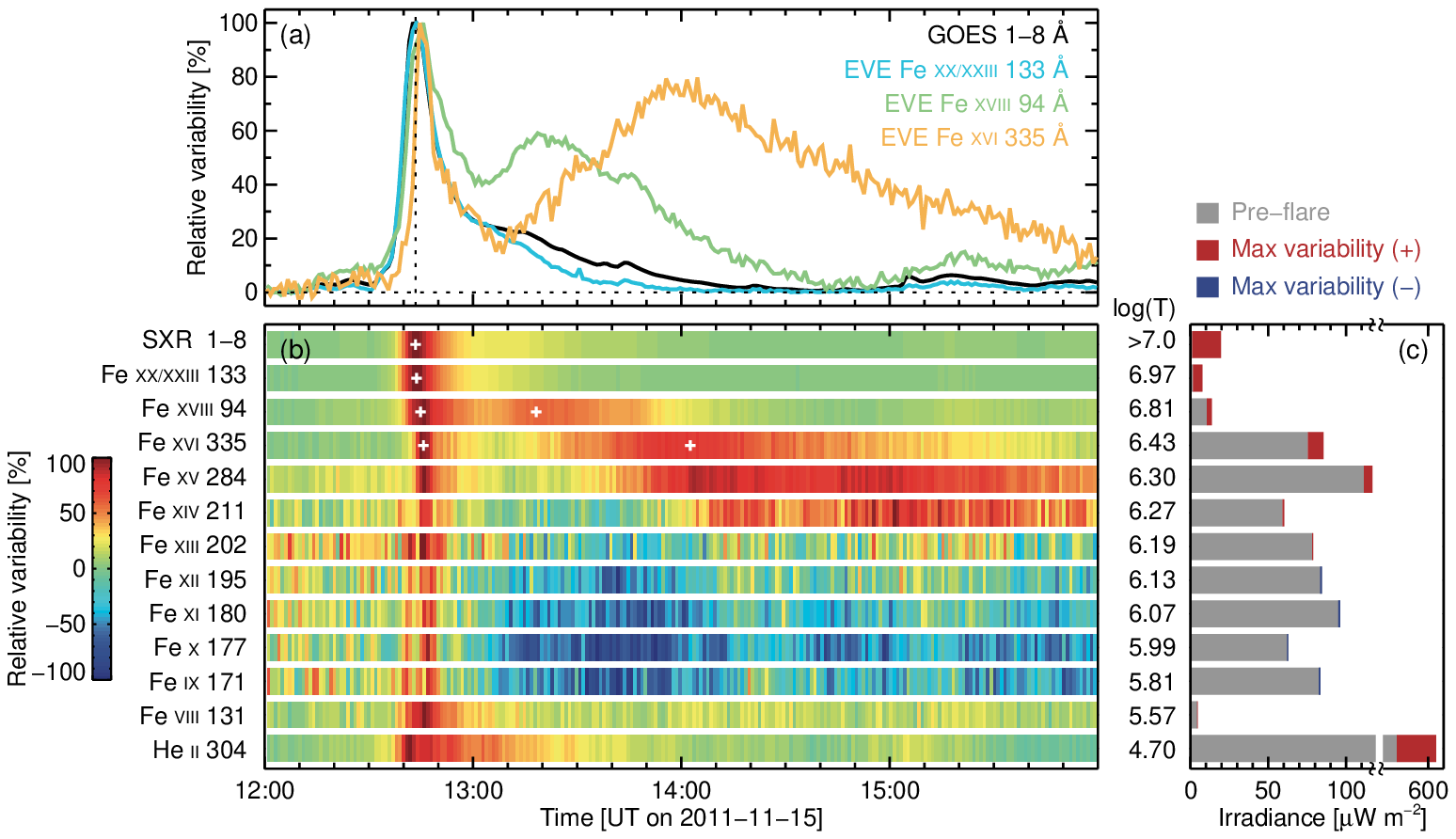}}
\caption{EVE flare irradiance measurement. (a) Relative irradiance variability in \textit{GOES} SXR, EVE \ion{Fe}{20}/\ion{Fe}{23} 133~{\AA}, \ion{Fe}{18} 94~{\AA}, and \ion{Fe}{16} 335~{\AA} lines. The minimum of five-minute running-averages between 12:00 and 12:30 in each line is subtracted as background. Values are then normalized by the maximum of the impulsive phase. The late phase shows up as a distinctive secondary peak in 94 and 335~{\AA}. (b) Relative irradiance variability for major emission lines shown in a dynamic spectra format, following \cite{woods2011}. In each row, emission increase is colored red, decrease (dimming) blue. In lines where dimming dominates (e.g. \ion{Fe}{10} 177~{\AA}), values are normalized by the absolute value of the maximum decrease. From top to bottom, lines are ordered based on their characteristic temperatures (listed on right) from hot to cool. White crosses show the irradiance peaks for some lines. (c) EUV irradiance in physical units. Pre-flare values are in gray. Red rightward extensions indicate the maximum increase; blue leftward extensions show the maximum decrease.}
\label{f:eve}
\end{figure*}


\begin{figure*}[t!]
\centerline{\includegraphics{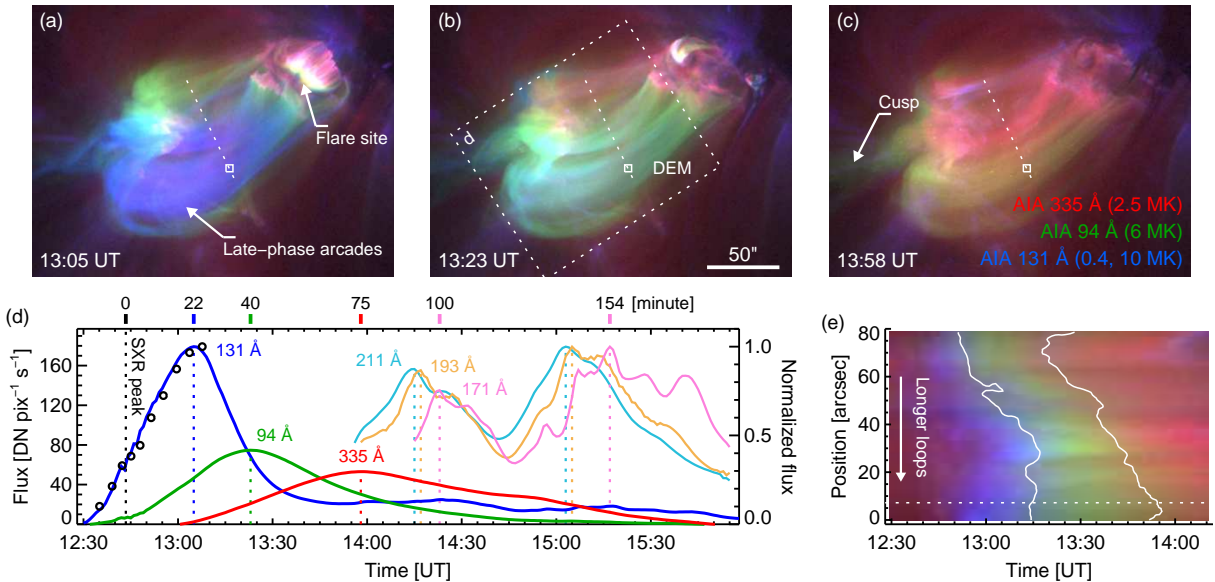}}
\caption{Late-phase arcades observed by AIA. (a)-(c) Tricolor AIA images during the flare decay phase. Emission in AIA 131~{\AA} (blue), 94~{\AA} (green), and 335~{\AA} (red) passbands peaks sequentially, resulting in different composite colors. A small segment of the loop (5$\times$5 pixels average, white square) is used for DEM analysis (Figure~\ref{f:dem}). A cusp-shaped structure is visible long after the eruption. (d) Light curve of the post-reconnection loops from SXR and EUV images. The EUV values are computed from the dotted box area in (b) with pre-flare values (12:00--12:20 UT) subtracted. For clarity, 211, 193, and 171~{\AA} values are normalized by their respective maximum; earlier dimming episodes in these passbands are not shown. Open circles show the SXR flux deduced from \textit{Hinode} XRT images with the Al-med filter, normalized to the maximum of the 131~{\AA} band. The peak times in EUV are indicated by vertical dotted lines. Their delay from the \textit{GOES} SXR peak are denoted on the top of the panel. (e) Space-time diagram constructed from the cut in (a)-(c). The changing colors illustrate a cooling process, with longer loops at the bottom changing more slowly. The white curves are contours of flux ratio between two AIA passbands (left, 94~{\AA} over 131~{\AA} at 0.4; right, 335~{\AA} over 94~{\AA} at 0.9). The horizontal dotted line denotes the pixel on which the DEM inversion is performed. (An animation of this figure is available online \href{http://sun.stanford.edu/~xudong/Article/Spine/cooling.mp4}{at this url}.)}
\label{f:cooling}
\end{figure*}


\begin{figure}[t]
\centerline{\includegraphics{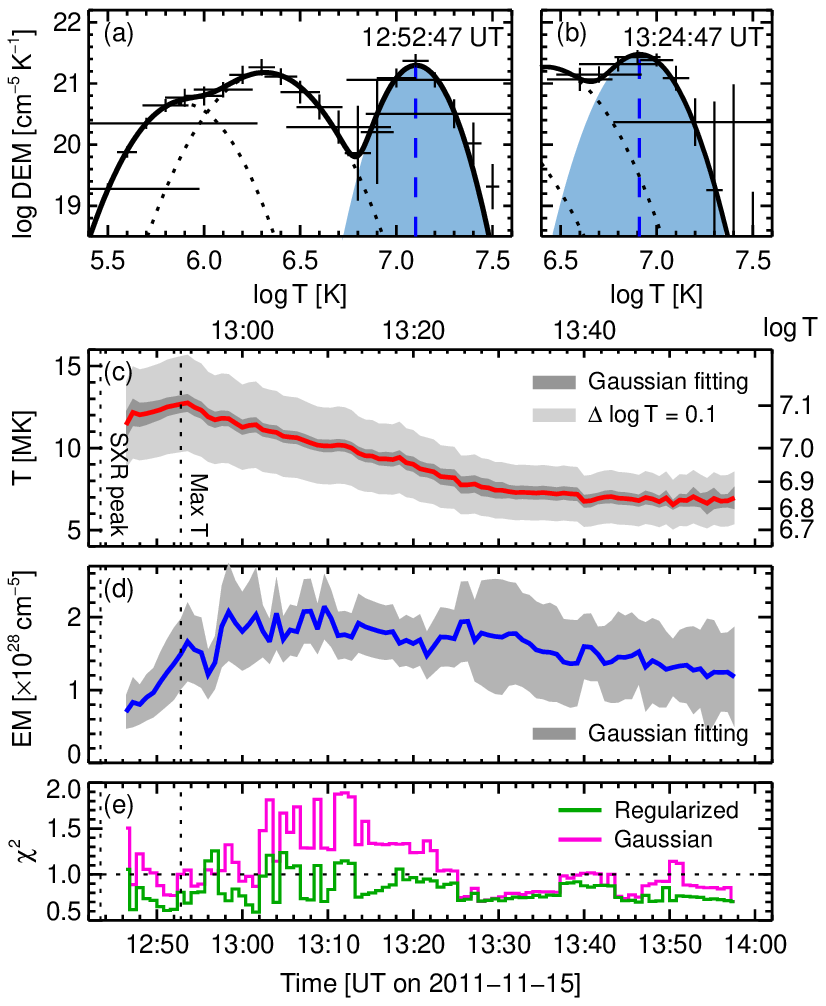}}
\caption{Evolution of thermal properties of a typical cooling late-phase arcade segment. (a) Regularized DEM inversion result at the inferred temperature maximum. A triple Gaussian model (curves) is used to fit the DEM profile (crosses) and to isolate the foreground hot loop contribution (cyan). A peak is found at $\log T\approx7.1$. (b) Similar to (a), for the hot component 32 minutes later. Peak temperature is now at $\log T\approx6.9$. (c) Evolution of loop temperature. The Gaussian centroid of a series of DEM solutions are plotted at a 48-s cadence. The times of SXR peak and temperature peak are marked by vertical dotted lines. The narrow, darker gray band is the 1$\sigma$ error from the fit. The lighter gray band show the $\pm0.1$ interval in $\log T$ as the estimated lower limit of the uncertainty in DEM analysis. (d) Evolution of the loop emission measure (EM). The gray band is from the $1\sigma$ fitting error. (e) Reduced chi-square ($\chi^2$) for the regularized DEM and the Gaussian model in data space. Larger $\chi^2$ of the Gaussian model between 13:00 and 13:20 UT stems from the bins around $\log T=6.3$ rather than the hot component.}
\label{f:dem}
\end{figure}


We now turn our attention to some of the consequences of the eruption. The eruption through two coupled magnetic systems yielded multiple groups of post-reconnection loops. All are discernible in SXR and 131~{\AA} images shortly after the flare peak (Figure~\ref{f:arcade}(b)). Similar to our toy model, the A1-type loops connected P1 and N1; they presumably originated from the progenitor sigmoidal flaring loops and appeared less sheared by this time. The A2 loops connected N2 and P2B (outside the eastern, or ``inner'' lobe of the fan), while the A3 loops connected N2 and P1/P1A (outside the western, or ``outer'' lobe of the fan). The connectivities of these loops are well represented in the toy model, but the loops here have more drastic contrast in size. We estimate the typical A1 loop to be about 30~Mm, A2 about 150~Mm. While the A1 loops appeared within a few minutes in all EUV passbands shortly after the eruption, the longer A2 and A3 loops showed up much later in the cooler passbands, delayed by 1--2 hours in 193~{\AA} (Figures~\ref{f:arcade}(g) through (i)). We link this to the EUV late phase in Section~\ref{sec:latephase}. It is interesting to note that the A2/A3 loops appeared in 193~{\AA} sequentially in a counterclockwise fashion. The slipping-type reconnection was perhaps present even during or after the eruption. 

Photospheric magnetic field displayed rapid, irreversible changes during the eruption. Comparing two relatively well observed frames (see Appendix~\ref{a:topo}) at 12:00 and 13:36 UT, the root mean square (RMS) of the horizontal component $B_h$ near the P1/N1 PIL increased by 18\% (Figures~\ref{f:topo}(a) and (b)); the distribution peak shifted from $\sim$800~G to $\sim$1000~G. The radial component $B_r$ was little affected. This is in line with our early study for a different event \citep{sun2012nlfff}, which appears to be a general feature of many major explosive events \citep{wangs2012}. We note that $B_h$ at the PIL kept increasing gradually even after the event, presumably due to flux cancellation.

Reconnection rearranges the field connectivity. We observe rapid penumbral decay on the outer (eastern) edge of N1 (blue star in Figure~\ref{f:topo}(a)), similar to previous reports \citep[e.g.][]{liuc2005}. This indicates a decrease of the longitudinal field ($B_l$), which corresponds to a decrease of $B_h$ since the AR was relatively far from the central meridian. The NLFFF extrapolation shows that the field lines originally connected N1 to P2A/P2B then connected to P1/P1A (blue in Figures~\ref{f:topo}(c) and (d)). The originally highly inclined loop legs straightened there, consistent with the observed photospheric field vectors.

The separatrix/QSL footprints are expected to change during the breakout process \citep{aulanier2000}. We compute the squashing degree $Q$ derived from the NLFFF model on the lower boundary. A signed logarithm version of $Q$, ${\rm{slog}}\,Q$ (Appendix~\ref{a:topo}), provides information for both the field mapping gradient and the field polarity \citep{titov2011}. Comparing the maps before and after the event (Figures~\ref{f:topo}(e) and (f)), we notice an inward shift of the fan footprint on the northern part, or, part of the flux ``moved'' out of the fan dome. Pre-explosion field lines (green) were once closed there, but became high-lying and connected to N2 afterwards (Figures~\ref{f:topo}(c) and (d)). This is consistent with the scenario that local flux is transported out of the fan dome through null-point reconnection. Conversely, the fan footprint shifted outward in the southwest. Flux ``moved'' into the fan dome, and the field lines (orange) became closed. This change is necessary as the overall flux under the fan (the total of N1) is conserved for a confined process.

The eruption likely altered the fan-spine structure. Its effect on the modeled QSL is not clear.

The modeled field lines generally appeared less sheared after the eruption, suggesting the magnetic helicity and free energy were reduced. Direct observation of the relaxed core field have been reported using XRT images \citep{suyn2007xrt}. The shear may have been transferred into the ejecta during the eruption \citep{aulanier2012}.


\section{The EUV Late Phase}
\label{sec:latephase}

EVE irradiance observations show a clear peak in many Fe lines early in the gradual phase (Figure~\ref{f:eve}(b)). The peak times lagged the \textit{GOES} SXR peak by a few minutes; lines with lower characteristic temperature (cooler) lagged slightly more. The \ion{He}{2} 304~{\AA} line peaked two minutes before the SXR, typical for flares with a strong impulsive phase \citep{woods2011}.

Emission from Fe lines cooler than $\log T\approx6.3$ dropped below the pre-flare level soon after the impulsive phase. This corresponds to the coronal dimming observed in AIA images. By inspecting the AIA difference images before and after the flare, we find the dimming regions are mainly localized near the footpoints of the A2/A3 loops, presumably caused by mass loss from the eruption. The magnitude of dimming exceeds the irradiance peak for a few lines (e.g. \ion{Fe}{10} 177~{\AA}). The absolute variation, nevertheless, is small compared to the pre-flare background (Figure~\ref{f:eve}(c)).

This event has a clear EUV late phase. No obvious emission increase appeared in the \textit{GOES} SXR or EVE \ion{Fe}{20}/\ion{Fe}{23} 133~{\AA} irradiance after the impulsive phase, a selection criterion for the late-phase event \citep{woods2011}. A secondary peak showed up in \ion{Fe}{18} 94~{\AA} and \ion{Fe}{16} 335~{\AA} (Figure~\ref{f:eve}(a)), delayed from the SXR peak by 35 and 81 minutes, respectively. Their variability was about 60\% and 80\% of the irradiance peak. In some cooler lines, a secondary peak appeared with further delay (e.g. \ion{Fe}{15} 284~{\AA}; Figure~\ref{f:eve}(b)); in others the dimming was reduced (e.g. \ion{Fe}{12} 195~{\AA}). This sequential delay in increasingly cooler lines is consistent with a cooling process of the post-reconnection loops.

Imaging observations show that the late-phase emission came mainly from the A2 and A3-type post-reconnection loops (Figure~\ref{f:cooling} and accompanying animation). For the XRT and AIA 131~{\AA} images, the flux count of the late-phase arcade displays a well-defined secondary peak 22 minutes after the \textit{GOES} maximum (Figure~\ref{f:cooling}(d)). The peak was relatively weak compared to the peak of the whole region and occurred early during the gradual phase. It thus became indistinguishable from the main flare peak in \textit{GOES} and EVE data. This demonstrates that in an EUV late-phase event, responsible loops can be heated to over 10~MK without being detected in the full-disk integrated irradiance observation.

AIA flux from the A2 and A3 loops peaked in the 94 and 335~{\AA} passbands 40 and 75 minutes respectively after the \textit{GOES} SXR (Figure~\ref{f:cooling}(d)), similar to the EVE observation. This secondary peak then appeared in 211, 193, and 171~{\AA}; the 171~{\AA} peak was delayed by about 100 minutes. An additional, third peak appeared another hour later (i.e. 171~{\AA} at 15:17 UT), which came from the even higher A3-type loops (e.g. Figure~\ref{f:arcade}(i)).

We now characterize the thermal evolution of these late-phase arcades. The composite AIA images from the 131, 94, and 335~{\AA} channels (Figures~\ref{f:cooling}(a)-(c)) show a similar loop morphology for an extended period of time. The color, or the relative intensity from different channels gradually changes, indicating a changing temperature. A vertical cut in the image samples the composite color of loops. The resulting space-time diagram shows a clear lag in color transition for longer loops towards the bottom, indicating longer cooling times for them (Figure~\ref{f:cooling}(e)). The initial cooling is likely dominated by conduction, since its time scale ($\tau_c$) varies with loop length squared ($L^2$), whereas the radiative cooling time scale ($\tau_r$) is independent of $L$ \citep[e.g.][]{reale2010}. This scaling law strongly differentiates loops of different lengths.

We focus on a small segment of what appeared to be A2-type arcades and perform a series of regularized DEM inversions using six AIA passbands \citep{hannah2012}. A new DEM peak indeed appeared above 10~MK (Figure~\ref{f:dem}(a)). Assuming the new peak comes from the foreground late-phase arcades, we fit the DEM curve with a triple Gaussian model to separate the hot component from the background (see Appendix~\ref{a:dem} for details). The fit is generally compatible with the data (Figure~\ref{f:dem}(e)).

The results here represent the average of the loop segments with similar temperature along the line of sight. They should perhaps be considered as an order-of-magnitude \textit{estimate} for the \textit{typical} late-phase arcades. Large systematic uncertainties that are not well understood exist in the coronal emission modeling \citep[e.g.][]{judge2010}, the ill-posed DEM inversion problem, and our ad hoc post-processing (Gaussian fitting) technique.

The loop temperature, estimated from the centroid of the hottest Gaussian component, reached its peak $T_0\approx13$~MK at 12:53 UT and then gradually cooled (Figure~\ref{f:dem}(c)). The time sequence of Gaussian has a stable, narrow width ($\sigma \approx 0.1$ in $\log T$), which translates to about 2--3~MK. This is also at the temperature resolution limit of the DEM technique \citep{judge2010}. Thus, it may be considered as a lower limit of the uncertainty. The temperature dropped to about 8~MK at around 13:25 UT (Figure~\ref{f:dem}(b)), then to about 7~MK at 13:58 UT. We stop the analysis at 13:58 UT when the arcades cooled enough that the hot component could no longer be consistently separated from the wide, cooler counterparts. 

The area below the hot component may be used as an estimate of the emission measure (EM), which is proportional to the density squared ($n^2$) integrated along the line of sight distance $h$ and multiplied by a filling factor $\eta$ (Appendix~\ref{a:dem}). At temperature maximum $T_0$, the EM (per unit area) is about $1.5\times10^{28}$~cm$^{-5}$ (Figure~\ref{f:dem}(d)). The maximum EM appeared later and is about $2\times10^{28}$~cm$^{-5}$; the peak time is quite uncertain given the large uncertainties in EM estimation. For a wide range of possible $h$'s between 1--100~Mm (assuming $\eta=1$), the initial density at temperature maximum $n_0$ adopts a value roughly between $10^9$ to $10^{10}$~cm$^{-3}$.


\begin{figure}[t]
\centerline{\includegraphics{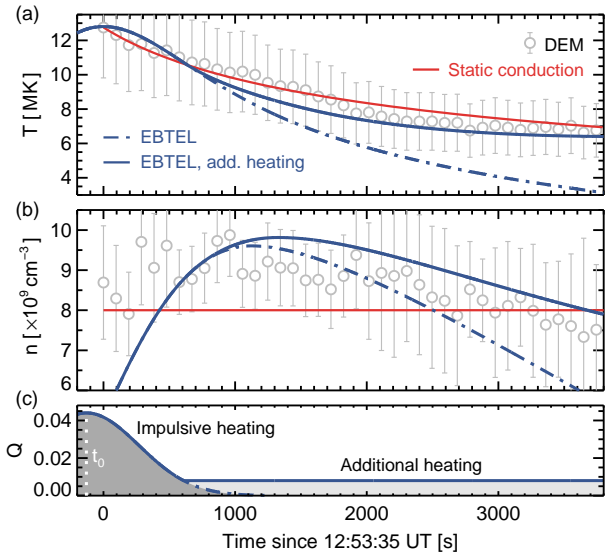}}
\caption{Modeled and observationally inferred properties of the late-phase arcade. (a) Temperature evolution. Open circles with error bars are from DEM at a 96-s cadence (taken from Figure~\ref{f:dem}). The red curve shows the static conductive cooling model. The dash-dotted blue curve shows the EBTEL model with impulsive heating only, whereas the solid blue curve includes additional heating. (b) Density evolution. A line-of-sight integration distance $h=2.2$~Mm and a filling factor $\eta=1$ is assumed to obtain $n$ from EM. (c) Heating function for the EBTEL model in~erg~cm$^{-3}$~s$^{-1}$. Dark gray shows part of the Gaussian shaped impulsive heating term, which starts from $-1330$~s. Light gray shows the additional heating. White dotted line indicates the maximum heating rate at $t_0=-130$ s. In all panels $t=0$ corresponds to temperature maximum at 12:53 UT; $t=3900$ s is 13:58 UT.}
\label{f:tcool}
\end{figure}


Can cooling explain the EUV late phase in this event, or is additional heating necessary? A definite answer perhaps requires modeling of a multitude of loops based on observations \citep[e.g.][]{qiu2012,liuwj2013,hock2012}. Here, we take a heuristic approach to discuss the question more qualitatively. We attempt to model the evolution of $T$ and $n$ under two simplifying assumptions. First, the DEM results reflect the average conditions of a single typical loop, or a collection of similar loops. Second, the loop is heated impulsively with no energy input late in the event. Additional heating is likely required if the modeled temperature cools significantly faster than the observations.

Given the loop parameters, conduction should dominate over radiation at beginning, i.e. $\tau_c < \tau_r$. If we ignore radiation and consider only a static conductive cooling process where the density is constant \citep{antiochos1976}, the modeled temperature evolution generally agrees well with the DEM result (Figure~\ref{f:tcool}(a); see Appendix~\ref{a:cooling} for details). However, the loop density is unlikely to be constant in reality \citep[Figure~\ref{f:dem}(d), see also][]{svestka1982}. Radiation should also contribute throughout.

To address these issues, we use an enthalpy-based thermal evolution loop (EBTEL) model \citep{klimchuk2008,cargill2012} that solves the coupled evolution of density and pressure. An impulsive energy input drives the loop evolution, which is represented by an ad hoc Gaussian heating term (Figure~\ref{f:tcool}(c)). We adjust the peak and the width of the Gaussian such that the resultant initial values $T_0$ and $n_0$ roughly match the DEM result. Many trial runs within the peak-width parameter space are performed (see Appendix~\ref{a:cooling} for details).

In contrast to the static cooling model, mass loss affects the temperature evolution in EBTEL. The decreasing density reduces the conductive cooling time scale ($\tau_c \propto n$); the inclusion of radiation further reduces the total cooling time. In all trial runs, the modeled $T$ falls faster than the observation after about 1000 s, when $n$ reaches maximum and starts to decrease (Figure~\ref{f:tcool}). All runs yield a temperature of $\sim$3~MK at 13:58 UT, much lower than the observed 7~MK.

This analysis constitutes a crude ``null test'' in which we disfavor the hypothesis. In the framework of the EBTEL model, additional heating appears to be required, in addition to the apparent long loop cooling time scale. As a test, we turn on a heating term a few tens of percent of the peak value late in the event. With this additional heating source, both $T$ and $n$ profiles better fit the observations (Figure~\ref{f:tcool}).

The additional heating may come from weaker reconnection higher in the corona \citep{woods2011}. Loops are successively heated, and a line-of-sight integration presumably yields the observed slower decay of $T$ and $n$. Note that in this case, the single-loop modeling above is not valid. In the SXR and EUV images, we do observe a cusp shaped structure long after the eruption (Figure~\ref{f:cooling}(c)), which is characteristic for ongoing reconnection behind the ejecta. Larger cusp structure also seems to exist above the A2 or A3 loops in EUVI-B observations (Figure~\ref{f:arcade}(e)).

We caution that the conclusion may be model dependent. Further testing with hydrodynamic simulation is thus worthwhile. The analysis here does not quantify the importance of additional heating relative to the energy loss. Nor does the conclusion necessarily apply to other EUV late phase events. Ensemble loop modeling and case surveys may help clarify these questions. 


\section{Concluding Remarks}
\label{sec:conclude}

We present a unique solar explosive event observed by \textit{SDO} that exhibits both the spine-fan topology and an EUV late phase. Analyses demonstrate an intimate relation between these two aspects, which is highlighted by the observed hot spine loops. The main findings are as follows.

\begin{enumerate}[parsep=0pt,partopsep=0pt,leftmargin=6mm]

\item Flux emergence in a pre-existing bipolar AR led to a largely closed fan-spine topology with non-potential structure under the fan separatrix. Strong shearing motion of the core field and flux cancellation likely initiated reconnection at the coronal null, creating a circular flare ribbon at the fan footprint and a remote brightening at the outer spine footprint. A newly reconnected, hot coronal loop bundle directly connected these two components. It presumably traced the evolving spine field line and part of the fan, providing direct evidence for the physical linkage between two distinct magnetic systems below and above the fan separatrix. The sequential brightening of the circular ribbon and the apparent shuffling motion within the hot loop bundle suggest that the slipping-type reconnection was also at work.

\item The null-point reconnection transferred flux across the fan separatrix, as shown by the growing volume of the hot loops. It effectively alleviated the immediate constraint on the core field, thus facilitated the subsequent eruption. This ``breakout'' process is relatively mild compared to the ensuing flare reconnection. Its relative importance (as opposed to, for example, plasma instability) in triggering the eruption is unclear. In this event, the horizontal photospheric field near the main PIL rapidly increased; modeling shows that the fan separatrix footprint shifted afterwards, and the field lines generally became less sheared. These aspects are expected for such a ``confined-to-eruptive'' event.

\item The secondary EUV irradiance peak came from the large overlying post-reconnection loops, which naturally formed in the eruption from such fan-spine structure. These arcades were heated to over 10~MK, but the corresponding irradiance signature was overwhelmed by the main flare site. The sequential delay of the late-phase peak in increasingly cooler lines suggests a cooling process. The large length of the arcades led to exceptionally long conductive cooling time. Additional heating may also contribute.

\end{enumerate}

We finally note that this event serves as a good example of cross-scale magnetic coupling. The two flux systems and their evolutional trends are closely linked by the fan-spine topology. Changes within the small close-field system beyond a critical point can lead to the eruption on a much larger scale.



\acknowledgments
We are grateful to T. Wiegelmann for the NLFFF extrapolation model, and J. A. Klimchuk for the EBTEL model. We thank J. Qiu, Q.-R. Chen, W. Liu, C.-L. Shen, A. Malanushenko, and K. Liu for discussion and help on the data analysis. X. Sun, J. T. Hoeksema, and Y. Liu are supported by NASA contract NAS5-02139 (HMI) to Stanford University. Y.-N. S. is supported by NASA contract SP02H1701R from LMSAL to SAO. The \textit{SDO} data are courtesy of NASA and the AIA, EVE, and HMI science teams. We acknowledge the use of \textit{GOES}, \textit{Hinode}, \textit{RHESSI}, and \textit{STEREO} data. Magnetic field lines are visualized with \texttt{VAPOR} (\url{http://www.vapor.ucar.edu}). The curve fitting is performed with \texttt{MPFIT} (\url{http://purl.com/net/mpfit}).

{\it Facilities:} \facility{\textit{SDO}}, \facility{\textit{STEREO}}, \facility{\textit{RHESSI}}, \facility{\textit{GOES}}, \facility{\textit{Hinode}}.



\appendix

\section{Construction and Topology of the Toy Model Field}
\label{a:model}

We place two dipoles below a Cartesian computation domain ($-256 \leq x,\,y \leq 256$, $0 \leq z \leq 512$): ${\bf{m}}_1=5.120\times10^6\times(\sqrt{2}/2,0,-\sqrt{2}/2)^{\rm{T}}$ at ${\bf{r}}_1=(96,0,-16)^{\rm{T}}$, ${\bf{m}}_2=1.024\times10^8\times(1,0,0)^{\rm{T}}$ at ${\bf{r}}_2=(0,0,-80)^{\rm{T}}$ (the superscript ``T'' denotes a transposition of vector). Together they yield a potential field (PF) at ${\bf{r}}$, that is, the sum of two dipole fields
\begin{equation}
{\bf{B}} = \sum_{i=1}^2 \frac{3({\bf{m}}_i \cdot \hat{{\bf{r}}}_i) \hat{{\bf{r}}}_i - {\bf{m}}_i}{r_i^3},
\label{eq:dipoles}
\end{equation}
where $r_i=|{\bf{r}}-{\bf{r}}_i|$, and $\hat{{\bf{r}}}_i=({\bf{r}}-{\bf{r}}_i)/r_i$. The more deeply rooted ${\bf{m}}_2$ emulates the pre-existing AR (P2/N2 in Figure~\ref{f:cartoon}). It creates the overlying, largely $-x$-directed background field. The shallow ${\bf{m}}_1$ represents the newly emerged AR (P1/N1) inside P2. On the lower boundary ($z=0$), $B_z$ ranges from a few hundred to about 2000 for P1/N1; about 100--200 for P2/N2. These values are comparable to that of a typical AR in Gauss.

We discretize the domain using a uniform grid size $d=1$ (513$^3$ grid points) for the construction of a NLFFF model. To describe the non-potential component of the newly emerged field, we introduce a fluxrope along the P1/N1 PIL low in the domain with axial flux $\Phi_a=4\times10^{20}$ and no poloidal flux. An iterative magneto-friction process relaxes the field towards a force-free state \citep{ballegooijen2004}. After 30000 iterations (typical for the procedure), we halt the relaxation and evaluate three metrics of the model for its quality: the mean Lorentz force $L_f$, the mean field divergence $L_d$, and the current weighted mean angle between the magnetic field and electric current $\sigma_j$ \citep[e.g.][]{schrijver2006}. For a 99$^3$-pixel sub-domain around P1/N1 ($37 \leq x \leq 135, -53 \leq y \leq 45, 3 \leq z \leq 101$), we find that $L_f=0.06$, $L_d=0.04$, and $\sigma_j=2^\circ.3$. Although still slowly converging, the field is sufficiently force-free for our demonstrative purpose \citep[cf.][]{metcalf2008}. The lowest few layers, which emulates the photosphere, are not force-free as the algorithm introduces a buoyancy force there during the relaxation process \cite[e.g.][]{bobra2008}. They are therefore excluded from the analysis. The fluxrope has relaxed to a sheared-arcade-like structure in the final state; its twist has significantly reduced. Strongly twisted or unstable final states can be achieved by increasing the axial flux \citep{suyn2011}.

The model field contains a single null point. For the PF case with analytical expression, we set Equation~\eqref{eq:dipoles} to zero and numerically solve the equation set. We find the null at ${\bf{r}}_p=(78.0444,0.0000,38.7580)^{\rm{T}}$. Alternatively, we apply a trilinear null-searching procedure on the discretized field \citep{haynes2007} and get ${\bf{r}}'_p=(78.0440,0.0000,38.7653)^{\rm{T}}$. The relative error amounts to $|{\bf{r}}'_p-{\bf{r}}_p|/d=0.7\%$, less than $1\%$ of the grid size. For the NLFFF model, the trilinear procedure find a null at ${\bf{r}}_n=(79.1237,-5.6192,43.5026)^{\rm{T}}$, close to but higher than the PF case \citep[e.g.][]{lugaz2011}. This is perhaps analogous to flux emergence on the Sun where the overlying field expands in response.

In this model, electric currents concentrate around the fluxrope. The spine and the fan, although away from the fluxrope, now also carry currents and become non-potential. We may calculate the direction of the spine and the fan surface from the eigenvectors (${\bf{v}}_i$) of the Jacobian matrix $M_{ij}=\partial B_i / \partial x_j$ ($i,j=1,2,3$) of the null ${\bf{r}}_n$ \citep[e.g.][]{parnell1996,sun2012topo}. Assuming the sub-grid field is trilinear, we may estimate the current density ${\bf{J}}$ near the null, which is non-zero. We find that current flows both along the spine and the fan (${\bf{J}}\cdot{\bf{v}}_3\neq 0$, ${\bf{J}}\cdot{\bf{v}}_1\neq0$). The spine is titled with respect to the fan plane normal by $\cos^{-1}(({\bf{v}}_1\times{\bf{v}}_2)\cdot{{\bf{v}}}_3)=11^\circ.6$, as opposed to the PF case where the two align.

We evaluate the squashing degree $Q$ at $z=3$ using an iterative scheme similar to \citet{aulanier2005}. The computation is performed for with a grid size $d_0=0.5$. For a grid point ${\bf{r}}_0$, we initially evaluate the field line connectivities from ${\bf{r}}_0\pm(d_0,0,0)^{\rm{T}}$ and ${\bf{r}}_0\pm(0,d_0,0)^{\rm{T}}$. This enables the evaluation of the field line mapping gradient $\partial X_i / \partial x_i$ ($i=1,2$), where $X_i$ is the coordinate of the conjugate footpoint ${\bf{r}}_1$ at $z=3$. The squashing degree $Q_0$ is then computed as the initial value using
\begin{equation}
Q=\dfrac{\sum\limits_{i,j=1}^2 \left( \dfrac {\partial X_i} {\partial x_j} \right)^2 }{B_{z,0}/B_{z,1}},
\label{eq:q}
\end{equation}
where $B_{z,0}$ and $B_{z,1}$ are the vertical field component at ${\bf{r}}_0$ and ${\bf{r}}_1$, respectively. The stepping is subsequently reduced by half, $d_1=d_0/2$, and the connectivities are evaluated again to derive $Q_1$ using Equation~\eqref{eq:q}. This iteration is halted at step $i$ when $Q_{i-1}/Q_i>0.99$, or $i>10$. Most pixels converge after a few steps, with $Q$ generally less than 10$^3$. For both the PF and the NLFFF, a single-pixel maximum with $Q\approx10^6$ appears in the ring-shaped footprint of the fan dome. With increasing resolution ($d_0$ decreasing), the peak $Q$ keeps increasing (over $10^{10}$ for $d_0=2^{-9}$), suggesting that the true separatrix footprint lies within.

The addition of the fluxrope adds substantial complexity to the field and distorts the fan dome. Such ingredient seems essential for an eruptive process to occur in this largely closed-field system. The global field morphology is similar for the PF and the NLFFF in this case, but can differ significantly in others, as discussed in our earlier study \citep{sun2012topo}. Caution should be exercised, since the application of a certain magnetic model is not known a priori.



\begin{figure}[t]
\centerline{\includegraphics{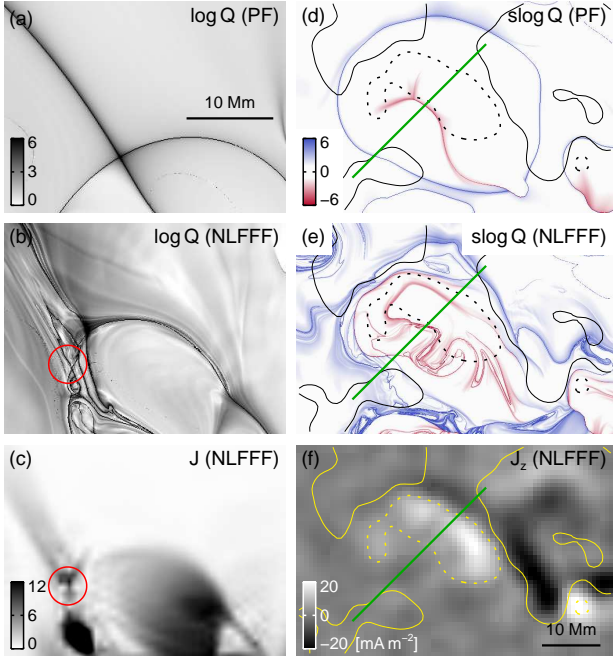}}
\caption{Squashing degree $Q$ and electric current density $J$ at the main flare site for the 12:00 UT frame. (a) Map of $\log Q$ on a vertical cross section through the PF model. (b) Same as (a), for the NLFFF model. The topology is less clearly defined. The weak field regions containing multiple nulls are denoted by the circle. (c) Map of $J$ on the cross section from the NLFFF model. (d) Map of ${\rm{slog}}\,Q$ on the lower boundary for the PF model. The circular footprint of the fan is apparent. (e) Same as (d), for the NLFFF model, identical to Figure~\ref{f:topo}(e). (f) Vertical current $J_z$ derived from the preprocessed boundary. In (d)-(f), the preprocessed vertical field is plotted as solid/dotted contours for $\pm$200~G. The green line segment shows the footprint of the vertical cross section. All $Q$ maps are computed at 8 times the field resolution.}
\label{f:qsl}
\end{figure}


\section{Magnetic Field Observation and Modeling for AR 11346 and 11351}
\label{a:topo}

The HMI pipeline derives a set of averaged Stokes parameters at a 12 minute cadence. Full disk vector field is inferred through a Milne-Eddington-based fitting algorithm, the Very Fast Inversion of the Stokes Vector \citep[VFISV;][]{borrero2011}. The 180-degree azimuthal ambiguity is resolved using a minimum energy method \citep{metcalf1994,leka2009}. Active regions are automatically identified, grouped, and extracted \citep{turmon2010}. Details and references to a series of papers describing the pipeline can be found in \cite{hoeksema2013}.

The vector maps are de-rotated and remapped using a Lambert equal area projection with a 0$^\circ$.06 resolution. This translates to about 1$\arcsec$, or 720 km at disk center. To fully cover the extended AR 11346, we extract a fairly large region of 512$^2$ pixels, equivalent to over 30$^\circ$ in Heliographic latitude and longitude. The field vectors at each pixel are decomposed into the zonal ($B_\phi$), meridional ($B_\theta$), and normal ($B_r$) component \citep{gary1990}. Those are used to approximate the $x$-, $-y$-, and $z$-component, respectively, on the planar lower boundary for our computational box. The height of the box is 256 pixels. Modeling results away from the center may be less accurate owing to the non-negligible curvature of the solar surface.

We use an optimization-based NLFFF extrapolation in Cartesian coordinate \citep{wiegelmann2004}. The algorithm aims to satisfy the following equations
\begin{eqnarray}
\nabla \times {\bf{B}} = \alpha {\bf{B}},\\
{\bf{B}} \cdot \nabla \alpha = 0,\\
{\bf{B}}(z=z_0) = {\bf{B}}_{\rm{obs}},
\label{eq:forcefree}
\end{eqnarray}
by minimizing the function
\begin{equation}
L=\int_V [B^{-2}|(\nabla \times {\bf{B}}) \times {\bf{B}}|^2 + |\nabla \cdot {\bf{B}}|^2] {\rm{d}}V.
\label{eq:minimization}
\end{equation}
Here $\alpha$ is the torsional parameter and varies in space, and ${\bf{B}}_{\rm{obs}}$ comes from the observed magnetogram. We assign a uniform altitude of $z_0=1$~Mm to the lower boundary, as the magnetogram is preprocessed to emulate the more force-free chromospheric boundary \citep{wiegelmann2006}. For comparison, we also perform a PF extrapolation using Green's function algorithm \citep{sakurai1989} with the same $B_z$ as boundary.

We model a total of 16 frames from 11:00 to 14:00 UT. Some of the frames contain known problems. For example, magnetic measurement at the flaring site during intense HXR emission might not be reliable since spectropolarimetric signals are susceptible to the changing thermal properties of the solar atmosphere \citep[e.g.][]{qiu2003}. In addition, the 180$^\circ$ azimuthal ambiguity resolution appears unreasonable in a small patch near the P1/N1 PIL in a few frames. This happens sometimes for complicated magnetic field observed away from the central meridian \citep[see][]{hoeksema2013}. The 12:00 and 13:36 frames are relatively well observed and close to the event, thus are chosen for detailed analysis.

From the extrapolation, we estimate the magnetic free energy $E_f$ by subtracting $\sum B^2/8\pi$ of the PF model from the NLFFF model. The value of $E_f$ decreased by about $2\times10^{31}$ erg from 12:00 to 13:36 UT, which is consistent with this low-M-class event. Nevertheless, the fluctuation is large in between. Owing to the poorly understood, large systematic uncertainties in modeling, the result should perhaps be regarded as an order-of-magnitude estimation \citep[e.g.][]{metcalf2008,sun2012nlfff}.

Using the trilinear method, we find a total of 8 computational cells containing null points above 8~Mm in the 12:00 UT NLFFF model. The mean height of these nulls is about 9~Mm; the RMS distance to their mean location is 3.1~Mm. The nature of these nulls varies; positive and negative, improper and spiral types coexist \citep[for classification, see e.g.][]{parnell1996}. The number, type, and exact locations change from frame to frame with no apparent pattern, but the mean location appears stable. Such behavior also appeared in our early study \citep{sun2012topo}. Owing to the imperfect boundary input and extrapolation algorithm, the residual Lorentz force and divergence are known to be unsatisfactory in the NLFFF extrapolation \citep{derosa2009}. For example, the current weighted mean angle between ${\bf{B}}$ and ${\bf{J}}$ is $\sigma_j=17^{\circ}.1$ (for the FOV of Figure~\ref{f:field}(a) below 47~Mm), compared to 2$^{\circ}$.3 for the toy model. As another example, we estimate the relative field divergence at the nulls using the proxy $|\sum \lambda_i|/\sqrt{\sum|\lambda_i|^2}$, where $\lambda_i$ $(i=1,2,3)$ are the eigenvalues of the Jacobian matrix at the null. The value should be zero for a strictly solenoidal field. For the NLFFF model, the minimum is 8\% (from a negative spiral null), much greater than the 1\% (negative improper null) of the PF model.

We note, however, that the separations between these derived nulls are small compared to the $\sim$30~Mm extent of the fan dome. These clustered nulls collectively define a weak field region. The global geometry and connectivity near this region prove to behave similarly as if near a true negative null point (Figures~\ref{f:field}(d) and (e)).

At the same time, the fan-spine topology is well defined in the PF model throughout. At 12:00 UT, the negative null is located at about 8~Mm, slightly lower than the NLFFF mean.

Following the procedures in \cite{pariat2012}, we evaluate $Q$ on a vertical cross section through the domain at the main flare site. The fan and spine are clearly present for the PF model (Figure~\ref{f:qsl}(a)), where $Q$ reaches over $10^{14}$ near their intersection. Similar structures exist for the NLFFF model (Figure~\ref{f:qsl}(b)), although less well defined and more complicated. The fan dome there is narrower and extends higher than the PF case. Current density in the NLFFF model is concentrated within the fan dome and along the spine (Figure~\ref{f:qsl}(c)). This explains the light twist in the modeled spine loop bundle (Figure~\ref{f:field}(e)).

On the lower boundary, we compute a signed logarithm version of $Q$, ${\rm{slog}}\,Q$, following \cite{titov2011}:
\begin{equation}
{\rm{slog}}\,Q = {\rm{sign}}(B_r)\,\log\,[Q/2+(Q^2/4-1)^{1/2}],
\label{eq:slog_q}
\end{equation}
where sign($B_r$) denotes the sign of the radial field component. It asymptotically approaches $\log Q$ at large values, while also preserves the information of the field polarity. Here, the fan footprint traverses the positive polarity field around N1 (Figures~\ref{f:qsl}(d) and (e)). The NLFFF $Q$ patterns appear much more complicated than the PF, owing to the non-potential structures in the model. The NLFFF fan footprint lies just outside the P1/N1 PIL on the northwest. Near the PIL, two narrow stripes of intense vertical current (Figure~\ref{f:qsl}(f)) coincide with high $Q$ values, which also resemble the morphology of the observed P1/N1 ribbons.


\section{DEM Analysis of the Post-Reconnection Loops}
\label{a:dem}

The thermal properties of optically thin plasma, along a line of sight distance $h$ with a volume filling factor of $\eta$, may be described by the differential emission measure (DEM)
\begin{equation}
\xi(T)={\rm{d}} (n^2 h \eta) / {\rm{d}}T,
\label{eq:dem}
\end{equation}
where $n$ is the plasma density at temperature $T$. The emission measure (EM) for a temperature range $T$ to $T+\Delta T$ is given by
\begin{equation}
{\rm{EM}}=\int_T^{T+\Delta T} \xi(T)\,{\rm{d}}T=n^2 h \eta.
\label{eq:em}
\end{equation}
For a set of AIA EUV observations, the DEM is related to the observed flux $g_i$ in the $i$-th passband
\begin{equation}
g_i=\int_T K_i(T)\,\xi(T)\,{\rm{d}}T+\delta g_i,
\label{eq:aiaflux}
\end{equation}
where $K_i(T)$ is the temperature response function, $\delta g_i$ the uncertainty.

Inversion from $g_i$ to $\xi(T)$ provides diagnostics of the plasma temperature and density. The problem is generally ill-posed, so additional constraint is required to obtain physically reasonable solutions. Here we employ a new regularized DEM inversion algorithm \citep{hannah2012,hannah2013} which have been tested favorably on a variety of theoretical and realistic cases. The scheme is robust and computationally fast. It also naturally provides both vertical (DEM) and horizontal (temperature resolution) error bars. The regularization term requires that the final reduced chi-square ($\chi^2$) is close to 1.

For this study, six AIA EUV passbands containing prominent Fe lines (94, 131, 171, 193, 211, and 335~{\AA}) are processed to Level-1.5 and coaligned to sub-pixel resolution using the standard \texttt{SolarSoft} modules. Only images with normal exposures (as opposed to the shorter flare mode) are used. Under this constraint, each set of six images are taken within 11 seconds. For spine loops with apparent slipping speed of about 100~km~s$^{-1}$ (Section~\ref{sec:eruption}), this leads to a 2.5-pixel displacement in two consecutive AIA images. We analyze the average of a 5$^2$-pixel region to partly mitigate this effect and to further reduce the noise.

We use the latest release of the AIA response function (version 4, February 2013) and an estimated uncertainty that is dominated by systematics, $\delta g_i\approx0.2g_i$ \citep{judge2010}. The temperature bin size is $\Delta \log T=0.1$. These choices sometimes yield large horizontal error bars for temperature bins between 7.0 and 7.3 (e.g. Figures~\ref{f:corona}(b) and \ref{f:dem}(a)), indicating the bins are not independent from each other. Larger bin size reduces the horizontal uncertainty \citep{hannah2012}, but are perhaps unfavorable due to our interest in the temperature evolution which has a small dynamic range (about 0.3 in $\log T$). Large vertical uncertainties in high temperature bins reflect the limitations of the DEM analysis when unknown systematic uncertainties are important.

The on-disk DEM inversion result contains contribution from the background. Direct subtraction of the pre-flare state fails in this case because the dimming in 171 and 193~{\AA} bands leads to negative relative flux. Another technique involves fitting a Gaussian to the flux profile along a single loop cross section and subtracting the base value \citep{aschwanden2011}. This is also difficult as there is no clean background pixels nearby for reference. As an ad hoc, post-processing solution, we fit a triple-Gaussian model to the regularized DEM \citep[cf.][]{aschwanden2011}
\begin{equation}
\xi=\sum \limits_{i=1}^3 \xi_i \exp \left[ - \frac{(\log T - \log T_i)^2}{2\sigma_i^2} \right],
\label{eq:tg}
\end{equation}
in the hope to separate the hot foreground component. The fitting consistently returns a hot, narrow component ($\log T_1$ at 6.8--7.1) well separated from the rest. Also obtained are a second, wider component ($\log T_2\approx6.3$), and a third cooler one ($\log T_3\approx6.0$). We use the first Gaussian as a proxy of the foreground loop DEM. The second component likely corresponds to the P2/N2 loops bright in 335~{\AA} in Figure~\ref{f:corona}(a). A double-Gaussian model yields almost identical results for the hot component, but does not describe the cooler components as well.

This procedure becomes invalid nearing 14:00 UT when the foreground loops cool and their signals blend with the rest. That is, the difference of two Gaussian centroid $T_1-T_2$ becomes similar to the sum of two half widths at half maximum, or $1.31(\sigma_1+\sigma_2)$. The fitting becomes inconsistent from one frame to the next. To estimate the ${\rm{EM}}$ and the density of the loops, we use the area between $\pm2\sigma$ below the Gaussian curve as a proxy. The integration is done at a $\Delta \log T=0.01$ step size.

We have neglected the horizontal error bars during the Gaussian fitting. For the post-reconnection loops, the Gaussian model has a similar $\chi^2$ as the regularized DEM except between 13:00 and 13:20 (Figure~\ref{f:dem}(e)). Our analysis is little effected because the larger $\chi^2$ mainly comes from a local peak at around $\log T=6.3$ that cannot be well described by a Gaussian profile (similar to Figure~\ref{f:corona}(b)). Jumps in the ${\rm{EM}}$ (Figure~\ref{f:dem}(d)) are partly caused by the small fluctuations in the fitted Gaussian width, that is, $\sigma_1$ ranging from 0.09 to 0.13.


\section{Cooling Time of Post-Reconnection Loops}
\label{a:cooling}


\subsection{Time Scales and the Static Conductive Cooling Model}
\label{subsec:tau}

We recount the estimation of loop cooling time scales and its application to a static conductive cooling model, mainly based on \citet{cargill1995}.

The cooling process in a coronal loop is governed by one-dimensional hydrodynamic energy conservation law. Specific forms of heat flux and radiation loss need to be prescribed. For example, with negligible flow speed and heating, one version writes
\begin{equation}
\frac{\partial{p}}{\partial{t}}=(\gamma - 1) \left[ \frac{\partial}{\partial{s}} \left( \kappa_0 T^{5/2} \frac{\partial{T}}{\partial{s}} \right) - n^2 \chi T^{-1/2} \right],
\label{eq:hd}
\end{equation}
where $p$, $n$, and $T$ denote pressure, number density of electron, and temperature, $t$ and $s$ denote time and loop location, respectively.

The first term on the right hand side is the conduction term, the coefficient $\kappa_0$ usually taken to be the classical Spitzer conductivity coefficient $\sim$10$^{-6}$ (in c.g.s. unit). The second is the radiation term in the form given by \citet{antiochos1980}, where $\chi$ is a constant, taken to be $1.2 \times 10^{-19}$ (in c.g.s. unit) here. For fully ionized hydrogen, we also have $\gamma=5/3$ and $p=2nk_BT$.

The characteristic cooling time of conduction $\tau_c$ and radiation $\tau_r$ can be estimated by equating the left hand side of Equation~\eqref{eq:hd} to each of the two terms on the right hand side. For example, we substitute $\partial{}/\partial{t}$ with $1/\tau_c$, and $\partial{}/\partial{s}$ with $2/L$, where $L$ is the \textit{full} loop length. After some rewriting, we get
\begin{equation}
\tau_c = 360\,n_{9}\,T_6^{-5/2}\,L_9^2\;\;{\rm{(s)}},
\label{eq:char_cond}
\end{equation}
\begin{equation}
\tau_r = 3450\,n_{9}^{-1}\,T_6^{3/2}\;\;{\rm{(s)}},
\label{eq:char_rad}
\end{equation}
where $n_{9}=n/10^{9}$~cm$^{-3}$, $T_6=T/10^6$~K, and $L_9=L/10^9$~cm \citep[cf.][]{reale2010}.

For flaring loops, $T$ is large, so generally $\tau_c \ll \tau_r$, i.e. conduction dominates the early cooling process. As the loops cool, $\tau_c$ increases and $\tau_r$ decreases. When $T$ drops to a critical value $T_*$, the two cooling times become equal, $\tau_{c*} = \tau_{r*}$. After that, radiation becomes dominant.

To obtain $T_*$, let Equations~\eqref{eq:char_cond} and \eqref{eq:char_rad} be equal. For a static conduction process with constant density \citep{antiochos1976}, it is easy to verify that
\begin{equation}
T_* = T_0 \left(\frac{\tau_{r0}}{\tau_{c0}} \right)^{-1/4},
\label{eq:temp_star}
\end{equation}
where the subscript 0 denote the initial values at the temperature maximum.

Furthermore, the temperature evolution of static conductive cooling is given by \cite{antiochos1976}
\begin{equation}
T(t) = T_0 \left( 1+\frac{t}{\tau_{c0}} \right)^{-2/5}.
\label{eq:stat}
\end{equation}
Thus, if radiation is negligible up to $T_*$, we can estimate the time $t_*$ needed to reach $T_*$ from Equations~\eqref{eq:temp_star} and \eqref{eq:stat}
\begin{equation}
t_* = \tau_{c0} \left[ \left( \frac{\tau_{r0}}{\tau_{c0}} \right)^{5/8} - 1 \right].
\label{eq:time_star}
\end{equation}


\renewcommand{\thefootnote}{\fnsymbol{footnote}}
\begin{table}[t]
\begin{center}
\caption{Estimated loop cooling time scales{$^\dag$}.\label{t:cooling}}
\begin{tabular}{ccc|ccc|cc}
\toprule
$L_9$ & $T_{0,6}$ & $n_{0,9}$ & $\tau_{c0,3}$ & $\tau_{r0,3}$ & $\tau_{r0}/\tau_{c0}$ & $t_{*,3}$ & $T_{*,6}$ \\[1mm]
\hline
\multirow{4}{*}{15} & \multirow{4}{*}{13} & 
             3 & 0.40 & 54 & 135 & 8.2 & 3.8 \\
&  & 5 & 0.66 & 32 & 48 & 6.9 & 4.9 \\
 &  & 8 & 1.1 & 20 & 18 & 5.6 & 6.2 \\
 &  & 10 & 1.3 & 16 & 12 & 5.0 & 7.0 \\
\bottomrule
\end{tabular}\\
\end{center}
\thefootnote{$^\dag$}{Static conduction is assumed. All values are in c.g.s. units, with $L_9=L/10^9$~cm, $\tau_{c0,3}=\tau_{c0}/10^3$ s, etc. The inputs are \textit{full} loop length $L$, initial (maximum) temperature $T_0$, and initial density $n_0$. Model results include initial conduction cooling time scale $\tau_{c0}$, initial radiation cooling time scale $\tau_{r0}$, total conduction cooling time $t_*$ at when $\tau_{c}=\tau_{r}$, and temperature $T_*$ at $t_*$.}
\end{table}
\renewcommand{\thefootnote}{\arabic{footnote}}


We have estimated the initial loop parameters for the A2-type arcades. In Table~\ref{t:cooling}, we list the model outputs for four possible $n_0$ values. In call cases, there is $\tau_{c0} \ll \tau_{r0}$, and $T_{*,6} \le 7$. This means that conduction is dominant and remains more important for the entire time period we study, where $T$ cools from 13 to 7~MK.

The A1 loops are very different. Their typical length is about 30~Mm, only $\sim$20\% of the late-phase arcades. We thus expect a much shorter cooling time scale. We use the \textit{RHESSI} HXR observations to deduce their thermal property.

For ten one-minute intervals between 12:42 and 12:53 UT, we fit the HXR spectra between 10 and 80 keV with a thermal and a broken power law component. The highest thermal temperature is $20.7\pm0.1$ MK. The corresponding, volume-integrated EM is $(9.6\pm0.3)\times10^{47}$ cm$^{-3}$. We use the area $A$ within the 50\%-maximum contour in Figure~\ref{f:corona}(e) as an estimate of the HXR source size. Assuming a spherical shape, we estimate the volume as $4A^{3/2}/(3\sqrt{\pi})$, about $1.4\times10^{26}$ cm$^3$. For a filling factor of 1, the density is about $8\times10^{10}$ cm$^{-3}$. The fit yields a median reduced $\chi^2$ of 1.50.

Using these initial values, we estimate the conductive cooling time scale for A1 loops to be $\tau_{0,3}=0.13$, significantly shorter than A2/A3. Meanwhile, the modeled static conductive cooling (Equation~\eqref{eq:stat}) is also too fast compared to the \textit{RHESSI} observation (not shown here). To resolve the discrepancy, conduction needs to be suppressed, or energy needs to be continuously injected from a loop-top source \citep{jiangyw2006}. Alternatively, if a small volume filling factor is assumed \citep{takahashi2000}, the derived density becomes very high. The model will then agree better with the observation.



\begin{figure*}[th]
\centerline{\includegraphics{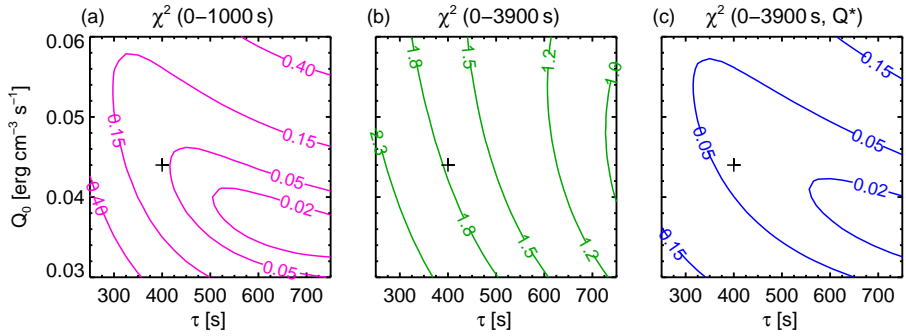}}
\caption{Parameter study of the energy input term (Equation~\eqref{eq:ebtel_heating}) for the EBTEL model. (a) Reduced chi-square $\chi^2$ between modeled $T$ and the DEM result for the first 1000~s. (b) Same as (a), for the whole 3900 s period. (c) Same as (b), for the whole 3900~s period but with additional heating. The values of $\chi^2$ are significantly reduced. The plus sign indicates the selected parameters used for Figure~\ref{f:tcool}, $Q_0=0.044$~erg~cm$^{-3}$~s$^{-1}$, $\tau=400$~s.}
\label{f:ebtel}
\end{figure*}


\subsection{Tests Using the EBTEL Model}
\label{subsec:ebtel}

The time-dependent, enthalpy-based thermal evolution loop (EBTEL) model \citep{klimchuk2008} calculates the mean properties of the loop plasma (i.e. zero-dimension) for specified energy input. It is efficient, and has been shown to agree well with more sophisticated one-dimensional hydrodynamic models.

We use a recently updated version \citep{cargill2012} that solves for density $n$ and pressure $p$,
\begin{equation}
\frac{dn}{dt} = -\frac{c_2}{5 c_3 k_B T} \left( \frac{F_0}{L} + c_1 n^2 \Lambda(T) \right),
\label{eq:ebtel_n}
\end{equation}
\begin{equation}
\frac{dp}{dt} = \frac{2}{3} \left[ Q(t) - (1+c_1) n^2 \Lambda(T) \right].
\label{eq:ebtel_p}
\end{equation}
Here, $F_0=-(2/7) \kappa_0 (T/c_2)^{7/2}/L$ (L the \textit{half} loop length) is the conductive flux at the base of the corona; $\Lambda(T)$ is the empirical radiative energy loss function \citep[see][]{klimchuk2008}; $Q(t)$ is the input heating term. The coefficient $c_1$ is determined self-consistently based on the loop parameters; $c_2$ and $c_3$ are fixed values determined from comparison with simulations. We have ignored the non-thermal beam heating term because the late-phase arcades did not show obvious HXR emission. Finally, the temperature is obtained as $T=p/2nk_B$.

We assume that the loops were heated impulsively early in the event. Following \cite{liuwj2013}, we adopt a Gaussian profile
\begin{equation}
Q(t) = Q_0 \exp \left[ -\frac{(t-t_0)^2}{2\tau^2} \right] + Q'.
\label{eq:ebtel_heating}
\end{equation}
We vary the free parameter pair $(Q_0,\tau)$, and let $t_0=3\tau$. After the modeling, the reference time is adjusted so that the maximum temperature appears at 12:53 UT, same as the observation. We require the final $t_0$ to fall between 12:43 UT (\textit{GOES} SXR flux peak) and 12:53 UT. An additional low background heating term $Q' \equiv 10^{-5}$~erg~cm$^{-3}$~s$^{-1}$ is used to obtain the initial equilibrium. Its low value has no impact on the evolution once the impulsive term sets in \citep{qiu2012}.

The modeled $T$ and $n$ are compared with the DEM values from 12:53 ($t=0$~s) to 13:58 UT ($t=3900$~s). For the parameter range $Q_0 \in [0.03,0.06]$~erg~cm$^{-3}$~s$^{-1}$ and $\tau \in [250,750]$~s, there is $T_0 \in (11.4,14.0)$~MK, and $n_0 \in (2.8,10.7)$~cm$^{-3}$. This roughly agrees with the DEM result. The temperature maximum usually appears 1--3 minutes after the heating maximum, and the density maximum about 1000~s after the temperature maximum.

A typical $T$ profile matches the DEM result at first, but soon is seen to decrease too fast (Figure~\ref{f:tcool}(a)). This is illustrated by the much larger $\chi^2$ for the whole period compared with the first 1000~s (Figures~\ref{f:ebtel}(a) and (b)). The final temperature at 3900~s is always around 3~MK. For this illustration, we use $Q_0=0.044$~erg~cm$^{-3}$~s$^{-1}$ so the modeled $T_0$ exactly agrees with DEM.

Comparison of $n$ involves an unknown value $h\eta$ in EM (Equation~\eqref{eq:em}). We choose $h\eta\approx2$~Mm so the resultant $n_0\approx8\times10^9$~cm$^{-3}$ agrees with the static conduction model that best fits $T$. Similarly, we adopt $\tau=400$~s for the EBTEL model so the maximum $n$ roughly agrees with the observation (Figure~\ref{f:tcool}(b)). Still, the modeled $n$ profile does not rise fast enough during the first 1000~s, and decreases too fast in the later stage. This is true for all $\tau$ values tested here.

To test the effect of the additional heating, we add to Equation~\eqref{eq:ebtel_heating} another term $Q^*$ such that $Q$ never drops below 0.008~erg~cm$^{-3}$~s$^{-1}$ (see Figure~\ref{f:tcool}(c)). This term, at its maximum, ranges between 13\%--27\% of the tested $Q_0$ values. With $Q^*$, the modeled $T$ agrees with observation better; the $\chi^2$ is reduced by 1--2 orders of magnitude (Figure~\ref{f:ebtel}(c)).

We note that our choice of the heating term is simplistic. The individual flaring loops are likely heated much more suddenly than what the slow-rising Gaussian profile indicates.


\end{CJK}



\bibliographystyle{apj}
\bibliography{spine}



\end{document}